\documentclass[11pt]{article}

\usepackage[margin=1in]{geometry}
\usepackage{mathpazo}
\usepackage[backend=biber,natbib=true,style=apa]{biblatex}
\usepackage{xltabular}
\usepackage{authblk}

\usepackage{microtype}
\usepackage{graphicx}
\usepackage{subcaption}
\usepackage{caption}
\usepackage{todonotes}

\usepackage{booktabs} 
\usepackage{hyperref}
\usepackage{bbm}
\usepackage[labelfont=bf]{caption}

\usepackage{titlesec}

\usepackage{psfrag}
\usepackage{amsmath}
\usepackage{amsfonts}
\usepackage{verbatim}
\usepackage{mathrsfs}
\usepackage{amssymb}
\usepackage{pifont}

\usepackage{multirow}
\usepackage{longtable}
\usepackage{listings}
\usepackage{tabularx}
\usepackage{subcaption}

\usepackage{color, colortbl}
\usepackage{xspace}
\usepackage{xr}
\makeatletter

\titleformat{\section}[block]{\normalfont\Large\bfseries}{}{0pt}{}
\titleformat{\subsection}[block]{\normalfont\large\bfseries}{}{0pt}{}
\titleformat{\subsubsection}[block]{\normalfont\normalsize\bfseries}{}{0pt}{}

\title{Whose ChatGPT? Unveiling Real-World Educational Inequalities Introduced by Large Language Models}
\author[1,2*]{Renzhe Yu}
\author[1]{Zhen Xu}
\author[3]{Sky CH-Wang}
\author[4,5]{Richard Arum}
\affil[1]{Teachers College, Columbia University}
\affil[2]{Data Science Institute, Columbia University}
\affil[3]{Department of Computer Science, Columbia University}
\affil[4]{School of Education, University of California, Irvine}
\affil[5]{Department of Sociology, University of California, Irvine}

\affil[*]{Correspondence: \href{mailto:renzheyu@tc.columbia.edu}{\texttt{renzheyu@tc.columbia.edu}}}

\date{\vspace{-5ex}}
\date{}
\begin{document}
\maketitle

\begin{abstract}
    The universal availability of ChatGPT and other similar tools since late 2022 has prompted tremendous public excitement and experimental effort about the potential of large language models (LLMs) to improve learning experience and outcomes, especially for learners from disadvantaged backgrounds. However, little research has systematically examined the real-world impacts of LLM availability on educational equity beyond theoretical projections and controlled studies of innovative LLM applications. To depict trends of post-LLM inequalities, we analyze 1,140,328 academic writing submissions from 16,791 college students across 2,391 courses between 2021 and 2024 at a public, minority-serving institution in the US. We find that students' overall writing quality gradually increased following the availability of LLMs and that the writing quality gaps between linguistically advantaged and disadvantaged students became increasingly narrower. However, this equitizing effect was more concentrated on students with higher socioeconomic status. These findings shed light on the digital divides in the era of LLMs and raise questions about the equity benefits of LLMs in early stages and highlight the need for researchers and practitioners on developing responsible practices to improve educational equity through LLMs.
\end{abstract}

\section{Introduction}
The release of ChatGPT in November 2022 for the first time exhibited the power of large language models (LLMs) to the general public and moved the discussions beyond computing research communities. Since then, many more LLMs have become publicly available and people from different sectors of the society have started to explore their potential to assist with work activities. With LLMs' abilities to achieve decent performance on cognitive tasks such as standardized tests~\citep{GPT4}, education becomes one of the most promising fields to be transformed by these technologies. LLMs can help alleviate the scarcity of educational resources and create more personalized learning experience through various applications, such as creating curricular material, providing tailored feedback, and facilitating discussion~\citep{Kasneci2023}. Prior to the release of ChatGPT, researchers at the intersection of AI and education were already developing and evaluating innovative ways of incorporating LLMs into teaching, learning and assessment activities~\citep{Yan2024}, and the availability of commercial LLM products boosted research along this line.

A focal issue surrounding educational application of LLMs is what they mean for closing existing disparities in opportunity, experience, and achievement. Advocates and many experimental studies thus far have argued that LLMs can close the gaps by creating tailored resources and experience for learners from underrepresented and disadvantaged backgrounds, and by helping educators reduce administrative tasks to focus more on individual students' needs. On the flip side, concerns arise that LLMs may exacerbate existing educational inequalities by creating new gaps and biases. For example, a precondition for benefiting from LLM applications is a certain level of technology and AI literacy, where underprivileged students already fall behind in the first place.

While LLMs are comparatively new, established research on mechanisms and consequences of digital inequalities can hint on what LLMs has brought to education in its early phase. The emergence of new technologies tends to create digital inequalities that mirror existing social and economic inequalities. Affluent, better-educated populations not only have more access to digital infrastructure, tools, and content (first-level divide) but are also more tech-savvy to take advantage of new technologies to maximize their utility (second-level divide). As a result, they may also reap more benefits from using the technologies compared to their less affluent peers (third-level divide)~\citep{vandijk2020}. In the age of artificial intelligence, these inequalities are still valid concerns but exist in more subtle forms as they interact with individuals and society in more diverse manners~\citep{Lutz2019,Carter2020}. Following the widespread adoption of LLMs, recent research has unveiled new forms of inequalities in many aspects of human society such as knowledge sharing, creative economy, and information seeking~\citep{Hui2024,Chanona2024,Daepp2024}. In educational contexts, most recent research has focused on experimenting with LLM innovations \textit{in controlled settings}~\citep{wang2024tutor,kumar2023math}. Much less work has documented LLM usage, outcomes, and gaps \textit{in the wild} except for those based on surveys of students and educators~\citep{Ruediger2024,Shaw2023}.

In this study, we examine micro-level educational inequalities in the age of LLMs. We focus on inequalities in academic writing because writing is a critical skill to develop and writing assistance has been one of the most common and established use cases of LLM tools in educational contexts. Based on large-scale writing records from authentic learning environments, we uncover how the emergence of LLM tools altered writing quality and its disparities across students from different linguistic and socioeconomic backgrounds. We find that as the public became more familiar with LLM tools, the writing gaps between linguistically disadvantaged and advantaged students were increasingly narrower. On the other hand, this promising change was somewhat more concentrated on high-SES students. These findings provide evidence on complicated patterns of digital inequalities incurred by LLMs and highlight the need for policy responses to carefully harness the power of LLMs towards an equitable, technology-empowered future of education.

\section{Results}

\begin{figure}[ht]
    \centering
    \includegraphics[width=0.6\textwidth]{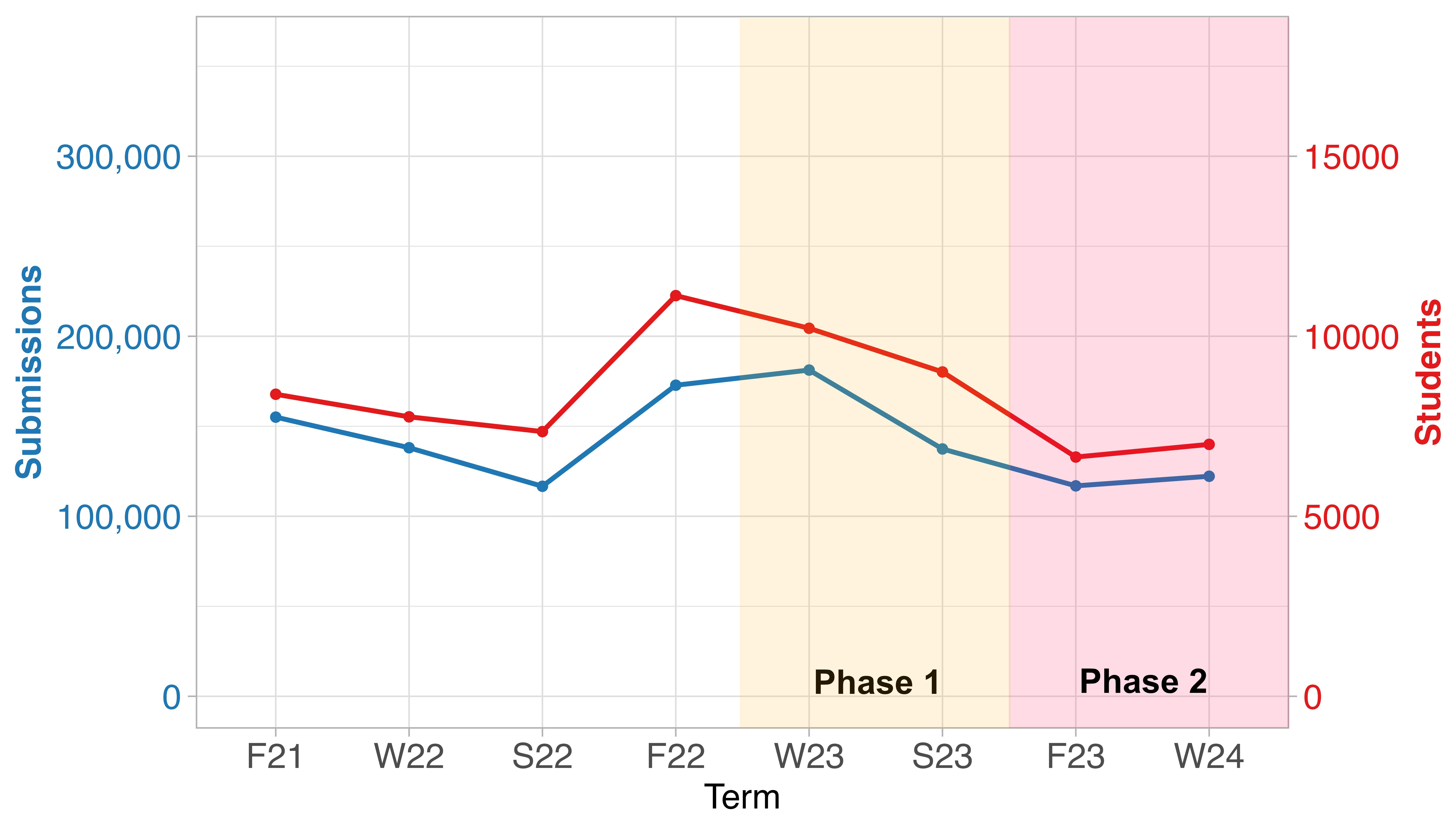}
    \caption{\textbf{Sample size over time.} Blue and red lines represent the unique number of forum submissions and students by academic term in the analytical sample. F: Fall (October to December); W: Winter (January to March); S: Spring (April to June).}\label{fig:sample}
\end{figure}

We analyzed 1,140,328 written submissions to forum-based assignments created by 16,791 undergraduate students across 2,391 courses between Fall 2021 and Winter 2024 terms at a four-year public university in the United States. LLM tools like ChatGPT were widely known and available at the end of Fall 2022, and we examined two phases of the LLM era: Phase 1 (January to June 2023) and Phase 2 (October 2023 to March 2024). Figure~\ref{fig:sample} depicts the sample size over time as well as the timing of the two phases.

\begin{figure}[ht]
    \centering
    \includegraphics[width=0.6\textwidth]{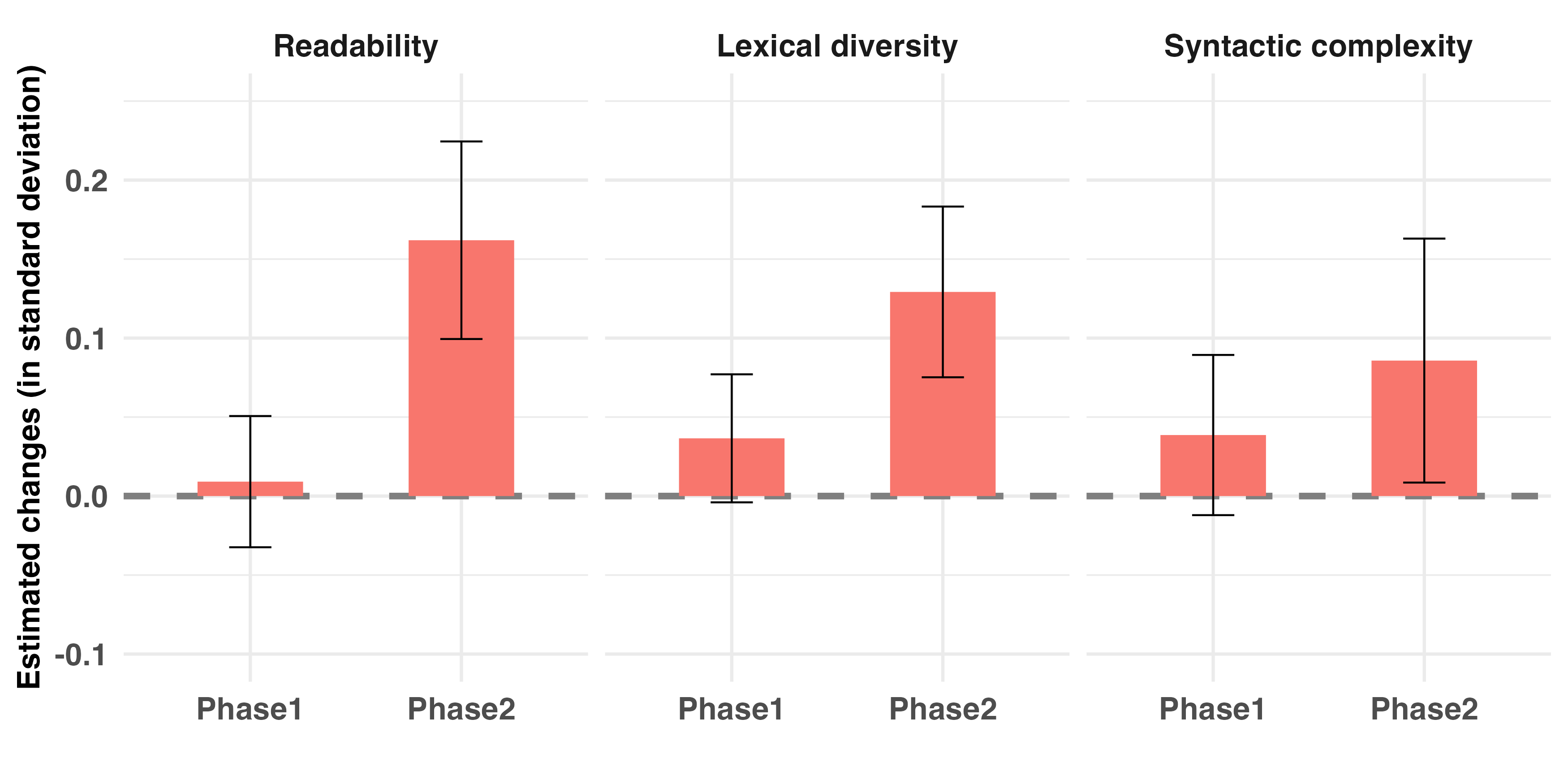}
    \caption{\textbf{Estimated changes in academic writing proficiency.} Each bar represents the point estimate of the average change (in standard deviation) in a composite writing proficiency index between a post-LLM phase and the pre-LLM period in the data. Phase 1: January 2023 to June 2023; Phase 2: October 2023 to March 2024. Error bars indicate 90\% confidence intervals.}\label{fig:rq1}
\end{figure}

We measured writing quality of forum submissions by three composite indices of language proficiency: readability, lexical diversity, and syntactic complexity. Each index was constructed by averaging the z-scores of two to three established linguistic measures to allow for comparisons across various models. With linear regression models, we first depicted the evolution of overall writing quality over time. Figure~\ref{fig:rq1} shows that across the three language proficiency indices, student writing quality was slightly higher (0.009 to 0.039 standard deviation) than the pre-LLM era in Phase 1 and substantially higher (0.086 to 0.162 standard deviation) in Phase 2, after controlling for various contextual and individual differences (detailed regression results in Table~\ref{tab:s1}).

\begin{figure}[ht]
    \centering
    \includegraphics[width=\textwidth]{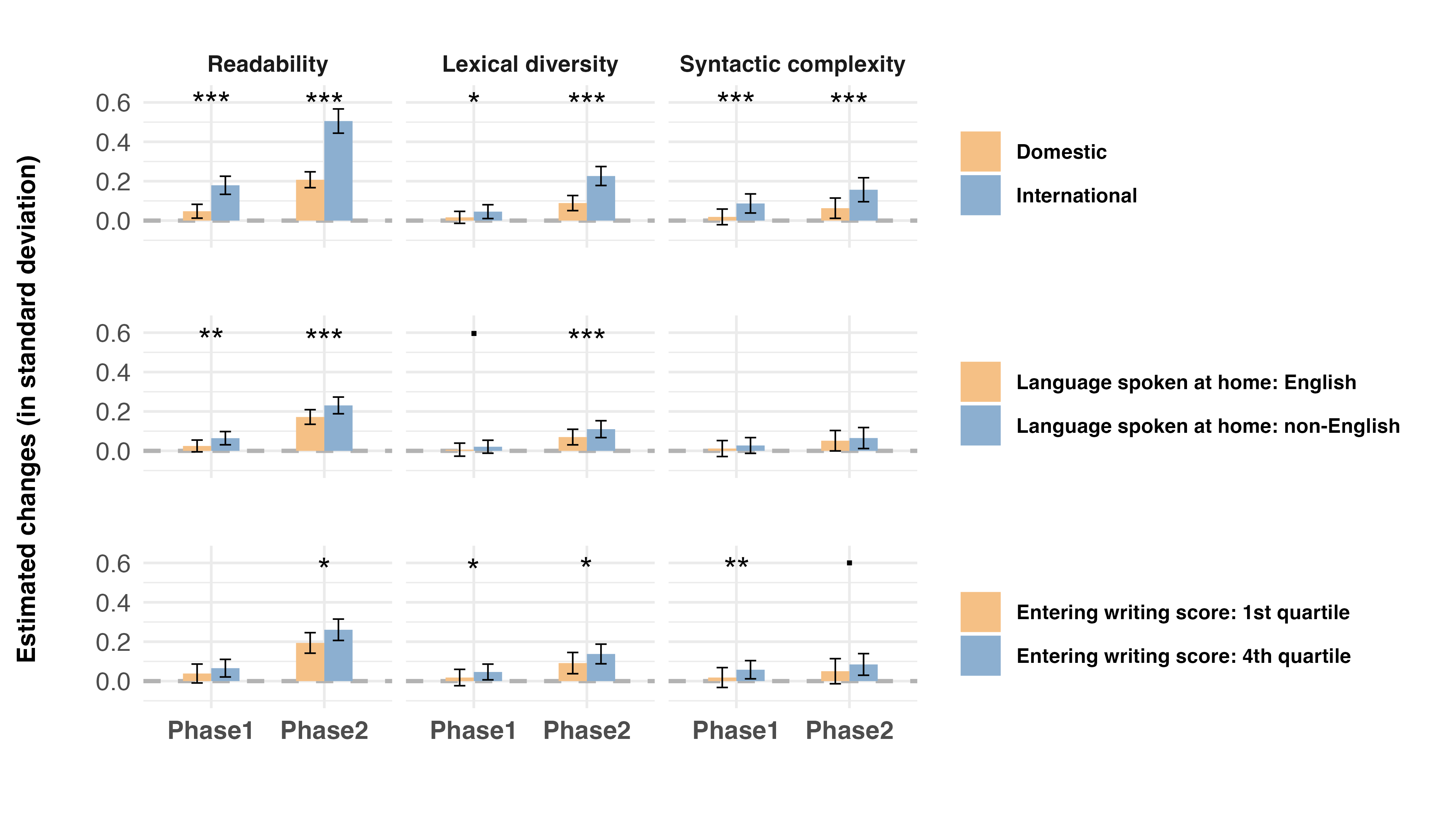}
    \caption{\textbf{Estimated changes in academic writing proficiency, by linguistic background.} Each bar shows the predicted average change (in standard deviation) in a composite writing proficiency index between a post-LLM phase and the pre-LLM period for a given student group. Phase 1: January 2023 to June 2023; Phase 2: October 2023 to March 2024. Error bars indicate 90\% confidence intervals. Statistical significance for the difference between adjacent bars (interaction term between the group and the phase indicators in the regression model) is denoted by asterisks: $p<0.10(\cdot)$, $p<0.05(^*)$, $p<0.01(^{**})$, $p<0.001(^{***})$.}
    \label{fig:rq2}
\end{figure}

More importantly, we were interested in whether the emergence of LLM tools leveled the playing field as many optimists had envisioned. We approached this question in two steps. First, as LLMs are especially capable of producing authentic language, we anticipated that LLMs could particularly aid students from linguistically disadvantaged backgrounds, helping to close the writing gaps between them and their more advantaged peers. Therefore, we leveraged linear regression models with interaction terms to examine how the writing of different groups changed differently after LLM tools became widely available (detailed regression results in Table~\ref{tab:s2}). Figure~\ref{fig:rq2} shows that while all student groups had somewhat more proficient writing in the LLM era, linguistically disadvantaged students (from non-English-speaking countries or families, or with lower entering writing scores) had significantly larger improvement than their peers across all indicators of writing language proficiency. Additionally, this differential improvement was increasingly pronounced over time. In Phase 1, linguistically disadvantaged students improved by 0.015 to 0.132 standard deviation (0.044 on average) more than their peers across various comparisons; in Phase 2, this differential improvement was as large as 0.014 to 0.298 standard deviation (0.088 on average). Given that linguistically disadvantaged student groups wrote 0.043 to 0.209 standard deviation less proficiently than their peers in the pre-LLM period, the estimated improvement patterns suggest that the emerging LLM tools helped close or even reverse language gaps in academic writing.

\begin{figure}[ht]
\captionsetup[subfigure]{justification=centering}
\centering
\begin{subfigure}{\textwidth}
\centering
\includegraphics[width=0.8\textwidth]{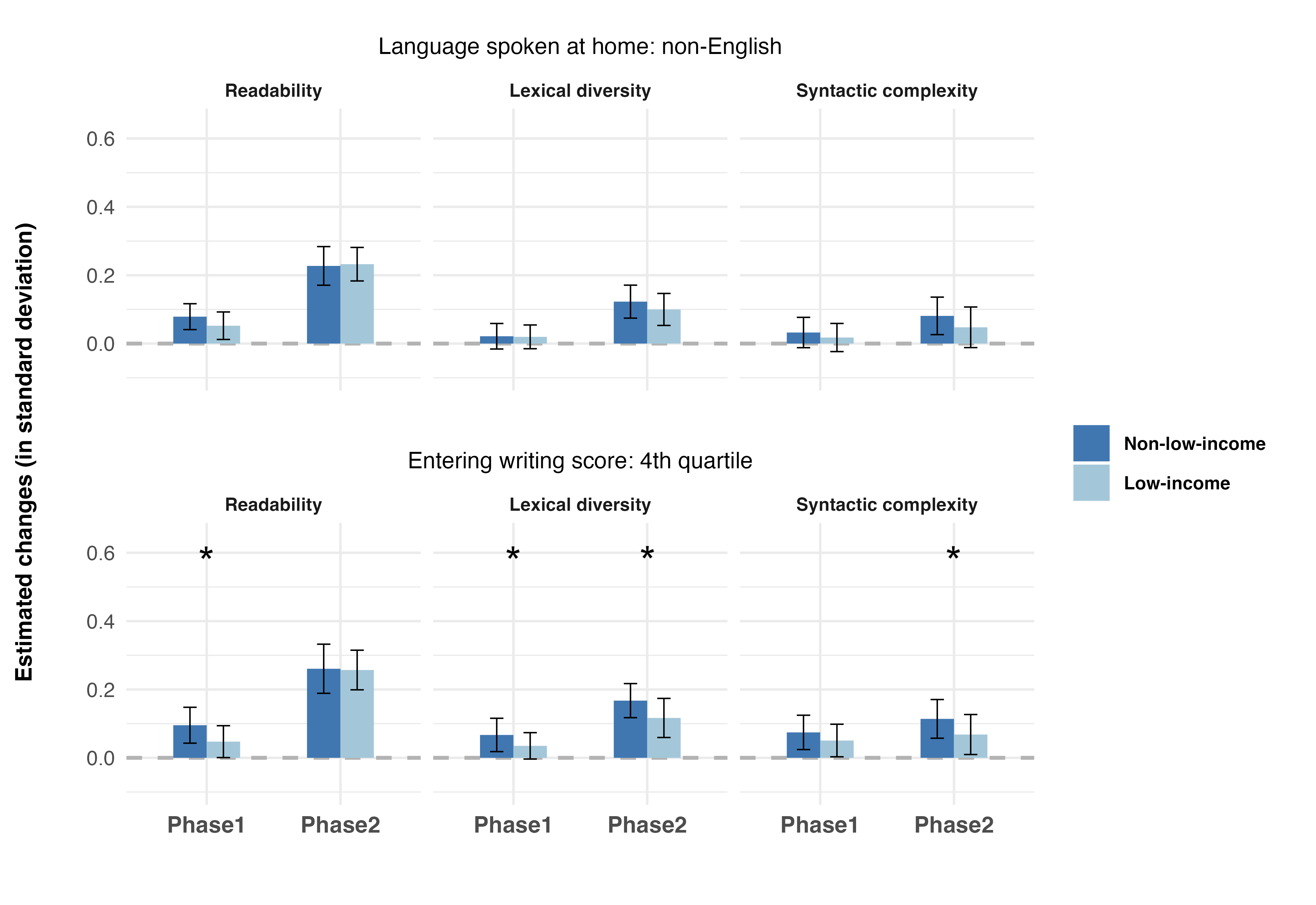}
\caption{SES: Family income}
\label{fig:rq3a}
\end{subfigure}
\begin{subfigure}{\textwidth}
\centering
\includegraphics[width=0.8\textwidth]{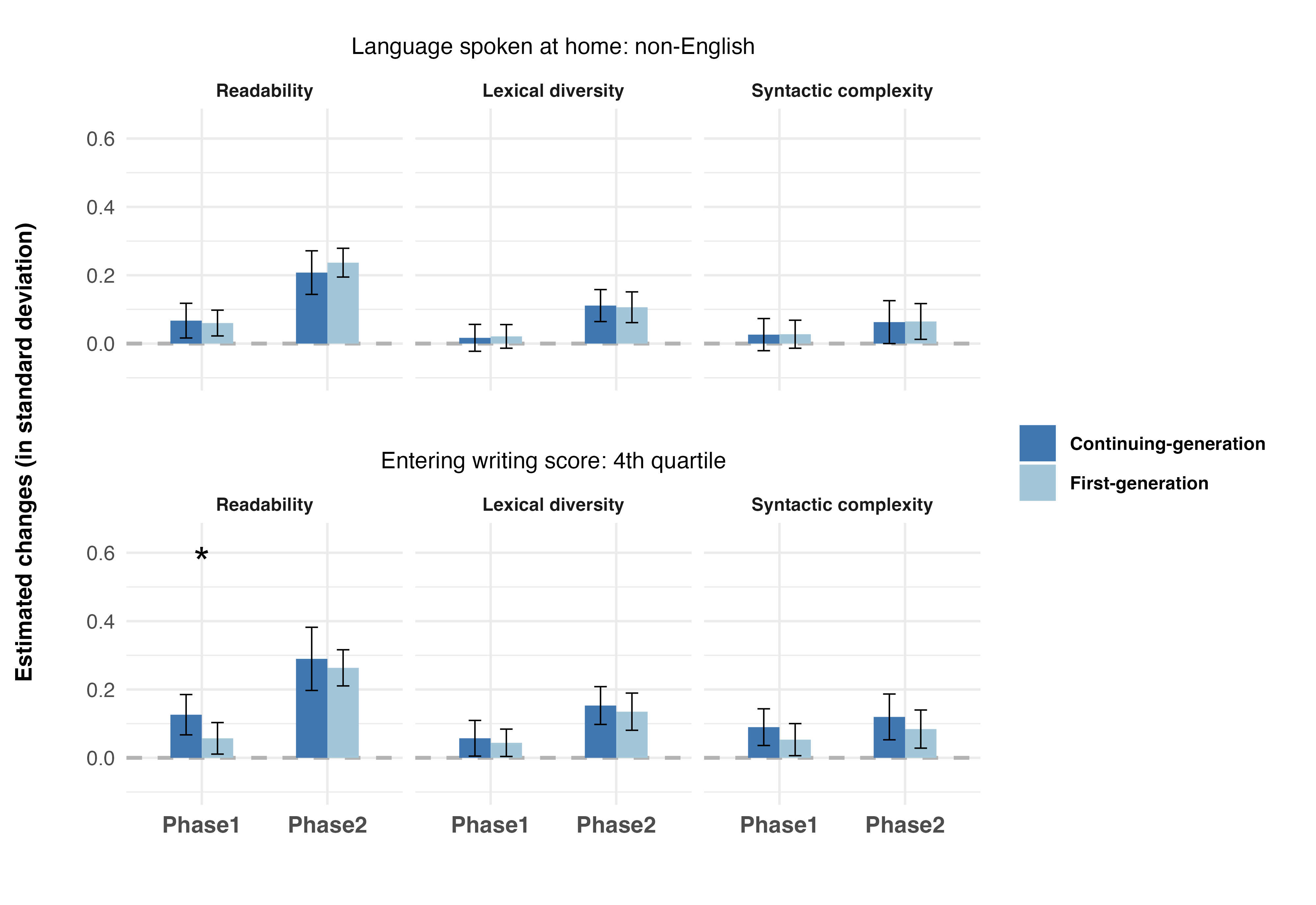}
\centering\caption{SES: Parental education}
\label{fig:rq3b}
\end{subfigure}
\caption{
\small
\textbf{Estimated changes in academic writing proficiency among linguistically disadvantaged student groups, by socioeconomic status (SES).} Each bar shows the predicted average change (in standard deviation) in a composite writing proficiency index between a post-LLM phase and the pre-LLM period for a given student subgroup. Phase 1: January 2023 to June 2023; Phase 2: October 2023 to March 2024. Error bars indicate 90\% confidence intervals. Statistical significance for the difference between adjacent bars (three-way interaction term between indicators of the linguistic group, socioeconomic group and phase in the regression model) is denoted by asterisks: $p<0.10(\cdot)$, $p<0.05(^*)$, $p<0.01(^{**})$, $p<0.001(^{***})$.}
\label{fig:rq3}
\end{figure}

While the narrowed writing proficiency gaps may suggest LLM tools' potential contributions to educational equity, we were cautious that these impacts may depend on students' socioeconomic status. As such, the second step of our equity investigation is examining students' writing changes broken down by both linguistic and socioeconomic backgrounds, by way of adding three-way interaction terms to the previous regression models (detailed regression results in Table~\ref{tab:s3}). Given that writing improvement was more prominent among linguistically disadvantaged student groups, we created Figure~\ref{fig:rq3} to highlight predicted socioeconomic differences only \textit{within} the disadvantaged groups. Across three language proficiency indices and two phases, the writing improvement attributed to LLM availability was mostly larger for high-SES than low-SES groups, with some estimates being more statistically significant than others. Students from low-income families on average experienced -0.005 to 0.051 standard deviation (0.025 on average) less writing improvement than their higher-income peers across various comparisons; this average gap was -0.029 to 0.069 standard deviation (0.015 on average) for students whose parents attended college compared to first-generation college students. These gaps existed in Phase 1 and persisted in Phase 2. In Figure~\ref{fig:s1}, we also illustrated the predicted socioeconomic differences \textit{within} linguistically advantaged groups, where the SES gaps were less consistent across linguistic indices and group definitions. These patterns combined provide some evidence of digital divides: the power of LLMs to equitize academic writing may be better leveraged by high-SES students, thereby overcoming language barriers at the cost of reinforcing social inequalities.

\subsection{Robustness check}
We performed a series of alternative analyses to validate the robustness of our findings. First, we used the same analytical models to examine changes in individual linguistic measures that composed the three composite indices. The findings were qualitatively similar to what was reported above (Tables~\ref{tab:s4}, \ref{tab:s5}, \ref{tab:s6}).

Second, we replicated the analyses on a subset of submissions to more ``important'' writing assignments, including those with greater grading variability and with longer content, because we assumed that students would take important assignments more seriously and be more motivated to use LLM tools if available. For submissions with varying grades, the results were comparable to our main findings (Figures~\ref{fig:s2}, \ref{fig:s3}, \ref{fig:s4}). For longer submissions, almost all the estimated gaps were larger than the main findings, reaffirming our claims about new digital inequalities introduced by LLM availability (Figures~\ref{fig:s5}, \ref{fig:s6}, \ref{fig:s7}).

Third, we expanded our sample to include submissions from as early as Fall 2019, enabling us to control for the impacts of the COVID-19 pandemic and compare recent writing patterns with pre-pandemic trends. The same analyses were then replicated. Most of the estimates were similar or more prominent in magnitude and statistical significance, strengthening the key takeaway messages (Figures~\ref{fig:s8}, \ref{fig:s9}, \ref{fig:s10}). For example, the average socioeconomic gaps in writing improvement were 0.028 standard deviation based on family income and 0.038 based on parental education, compared to 0.025 and 0.015 in the main results.

\section{Discussion}

We present a nuanced empirical analysis of real-world educational experience in the face of LLM availability. Focusing on college-level academic writing which was highly likely to be affected by the usage of LLM tools like ChatGPT, we find that the overall language proficiency in student writing gradually improved over time as LLMs became more familiar to the public, and that this improvement was increasingly more prominent for students from linguistically disadvantaged backgrounds, suggesting that having LLM tools available may have helped alleviate their existing language barriers. This aligns with the emerging research on the potential of LLMs to empower disadvantaged learners especially in language-related tasks such as second language writing~\citep{warschauer2023}.

On the other hand, our closer examination reveals that the equity benefits were not uniformly distributed. Specifically, the closed language proficiency gaps were slightly more concentrated among students from high-SES backgrounds. This gap persisted from the initial phase, when most people was new to LLMs, through the period of adjustment and training. As a result, it may have done little to or even exacerbated existing achievement gaps that LLM tools are expected to bridge. These observations are not surprising given the widespread digital divides introduced by previous generations of technologies~\citep{vandijk2020}, but such unintended consequences have been overshadowed by the early excitement about LLM innovation.

It is important to clarify that we did not estimate the actual usage of LLM tools as some recent studies have done \citep{Kobak2024,Liang2024}. Rather, we emphasize that it is the observed output that directly influences downstream consequences. In our context, instructors evaluate students' submitted writing regardless whether and how LLM tools were used during the creation process. Any identified writing gaps may directly translate into gaps in performance metrics which have more significant consequences in students' academic trajectories. We were not able to observe significant changes in grade performance in this study, because forum-based writing typically contributes only marginally to final course grades and is not usually assessed based on writing quality. However, because even these ``less important'' writing tasks revealed unintended inequalities, we expect more prominent disparities in higher-stakes writing artifacts.

Our study highlights the importance of researching novel technologies like LLMs \textit{in the wild} in addition to investigating innovations in lab or controlled environments. Such authentic evaluations are crucial for understanding the multifaceted social and institutional factors that may complicate the actual impacts of LLMs on day-to-day educational activities and outcomes. For one thing, learners who participate in early experimentation of the new technologies may feel excited about them and exhibit higher levels of motivation and productivity, which may not persist when the technologies become routine. In addition, most controlled studies are built upon access and consent, so there may be populations whose voices and needs are critical but systematically left out from research and development efforts around these new technologies. As LLM applications become integrated into the education system, research on digital inequalities needs to advance to capture more subtleties in access, literacies, engagement, and outcomes across a wider variety of educational activities.

While this study revealed nuanced digital inequalities through the lens of outcomes, risks of exacerbating existing inequalities can manifest at every phase of the LLM lifecycle in more complicated ways~\citep{Lee2024}. In this context, we advocate for a balanced approach between innovation and responsibility. Developers in pursuit of novel LLM applications should embed ethical considerations into the development process from the beginning, addressing issues like stereotypes reflected in the training data. Educators should closely monitor the real-world impacts of LLM applications, especially for learners from already disadvantaged backgrounds. Policymakers need to establish context-specific guidelines for equitable and accountable deployment of LLMs. These efforts should involve not only reactive measures but also proactive collaboration between different stakeholders and diverse learner communities to ensure that the benefits of LLMs are shared equitably, regardless of socioeconomic status, geographic location, or cultural background.

\section{Materials and Methods}
\subsection{Data and Sample}
Through an existing research partnership approved by the Institutional Review Board~\citep{arum2021}, we gained access to deidentified discussion forum data from the learning management system (LMS) deployed at a four-year public university in the United States. The LMS has been the major platform for course management across campus since Fall 2016, and discussion forum data captures authentic instructional and learning activities in real classes. While discussion forums can be used for various purposes across instructional contexts, we focused on forum-based graded assignments, which typically involve reflective writing or idea exchange related to the subject matter of the course~\citep{andresen2009}. Students' written submissions to these assignments can therefore reasonably capture their serious academic writing. The university runs on a quarter system, where each academic year (excluding summer) consists of three 10-week terms (quarters): Fall, Winter, and Spring. We retrieved the content and metadata of all forum-based submissions until March 2024, in which Winter 2024 term concluded. We defined two phases following the availability of LLM tools: Phase 1 (January to June 2023), representing the initial stage where most people were still exploring the new technology, and Phase 2 (October 2023 to March 2024), the period after a summer of adjustment and training.

From the university's administrative records, we also got deidentified information about students' linguistic and socioeconomic backgrounds, as well as students' course-taking records. Linguistic backgrounds included three variables as proxies for students' baseline English language proficiency: international student status, language spoken at home, and entering writing score. Most international students at the institution come from non-English-speaking countries, so this indicator can reasonably reflect students' English language proficiency. Entering writing score is a composite score of a student's entering writing skills that the institution calculates based on their relevant standardized test and AP scores as well as their performance in the institution's writing placement test before they enroll in any classes. Socioeconomic backgrounds included students' parental education and family income levels. Course-taking records documented each course a student completed in each semester and the letter grade they received, if applicable. We were able to join administrative records and discussion forum data via anonymized student and course identifiers.

The first prevalent LLM tool (ChatGPT) was released towards the end of Fall 2022. For our main analysis, we first kept written submissions between Fall 2021 and Winter 2024 (inclusive), allowing us to compare academic writing 4 quarters before and 4 quarters after LLM availability. We did not include earlier data to avoid the academic terms substantially disrupted by the COVID-19 pandemic. Then, we identified students who had any submissions both before and after LLM availability, and only kept their submissions, allowing for within-student comparison in the regression analyses. Finally, we removed submissions that were non-English and short (less than 10 words), that came from small courses (with fewer than 10 enrolled students), and that were created by students who did not complete the corresponding course.

With these sample restrictions, our main analytical sample included 1,140,328 posts created by 16,791 undergraduate students across 2,391 courses offered over the 8 terms. Of these students, 14.8\% were international, 24.0\% did not speak English at home, 37.9\% were first-generation college students, and 28.5\% came from low-income families.

\subsection{Measuring Academic Writing Quality}
We measured the quality of students' academic writing by quantifying \textit{language proficiency}, as early versions of LLM tools were particularly known for generating fluent English language and expected to alter the language proficiency of students' writing if actively used. Following established research in automated writing evaluation~\citep{Ke2019}, we included seven variables that capture three dimensions of language proficiency:
\begin{itemize}
    \item \textbf{Readability} is the ease with which a reader can understand a written text. We employed three standard lexical measures of a text's readability: Automated Readability Index (ARI), a characters-per-word measure~\citep{senter1967}; Dale-Chall readability formula (Dale-Chall), an adjusted measure based on a lexicon of ``difficult'' words~\citep{dale1948}; and Flesch-Kincaid grade level (FK Grade), a words-per-sentence and average syllables-per-word measure~\citep{flesch1948}.
    
    \item \textbf{Lexical diversity} corresponds to the range of different words used in a text~\citep{engber1995}. We used Jarvis's Measure of Textual Lexical Diversity (MTLD)~\citep{jarvis2013} and Carroll's corrected Type-Token-Ratio (CTTR)~\citep{carroll1964} as our primary measures.
    
    \item \textbf{Syntactic complexity} measures the sophistication and variety of syntactic forms in a text~\citep{Crossley2020}. We chose two measures of syntactic complexity: the max depth of the dependency tree (Max Depth) and the number of clauses (Clause)~\citep{chen2013}.
    
\end{itemize}

We computed the seven measures for each forum submission. For each measure, we ranked the values across all submissions in the analytical sample and removed the top and bottom 0.1\% to eliminate potential outliers. We then standardized the remaining values by computing z-scores. Finally, we separately averaged the z-scored measures in each foregoing dimension, forming composite indices of readability, lexical diversity, and syntactic complexity for each submission.

\subsection{Estimating Changes in Academic Writing Quality}
We leveraged linear regression models with fixed effects to estimate how overall writing quality changed following the release of LLM tools. We created two indicators for the two post-LLM phases, respectively ($\mathit{Phase_1}$ and $\mathit{Phase_2}$). Each indicator was equal to 1 for submissions created in the corresponding phase and 0 otherwise. We regressed each of the three composite indices on both indicators:

\begin{equation}\label{eq1}
    Y_{ijkt} = \alpha + \sum_{m=1}^{2} \beta_{1m} Phase_{m,t} + \sum_{n=1}^{27}\phi_{n} Topic_{n,i} + \lambda_j + \theta_k + \mu_{jt} + \epsilon_{ijkt}
\end{equation}
where $Y_{ijkt}$ is a composite index for submission $i$ created by student $j$ in course $k$ in term $t$. To control for differences in writing assignments, we applied Latent Dirichlet Allocation~\citep{blei2003} to uncover latent topics across all assignments and chose the optimal topic number $N=27$ with the largest coherence score~\citep{roder2015}. For each submission $i$, we added the weight of its corresponding writing assignment on the $n$-th topic, $Topic_n$, to the model. Finally, we included an array of fixed effects at student ($\lambda_j$), course ($\theta_k$), and year of study ($\mu_{jt}$) levels to adjust for individual, contextual, and developmental differences. The estimated $\hat{\beta}_{1m}$ captured the average change in writing proficiency in each of the phases, compared to pre-LLM times.

To unveil the equity implications of LLM tool availability, we first estimated how their emergence changed the writing gaps between students from different linguistic backgrounds. We defined three pairs of linguistically advantaged and disadvantaged student groups: domestic vs. international students; students who speak English vs. other languages at home; students in the top vs. bottom quartile of entering writing scores. We replaced student fixed effects in Model~\ref{eq1} with an indicator of linguistically disadvantaged group ($LingDis$) plus its interaction with each of the two phases. This was done separately for each linguistically disadvantaged group. The model is given below:

\begin{multline}\label{eq2}
    Y_{ijkt} = \alpha + \sum_{m=1}^{2} (\beta_{1m} Phase_{m,t} + \beta_{2m} Phase_{m,t} \ast LingDis_j) + \beta_3 LingDis_j \\+ \sum_{n=1}^{27}\phi_{n} Topic_{n,i} + \theta_k + \mu_{jt} + \epsilon_{ijkt}
\end{multline}
where the estimated $\hat{\beta}_{2m}$ captured the differential change in writing quality between the two groups following the availability of LLM tools. 

We further explored whether this differential change was conditional on students' socioeconomic status (SES). We identified two pairs of SES groups: students from non-low-income vs. low-income families; continuing-generation vs. first-generation college students. We added to Model~\ref{eq2} an indicator of low-SES group ($LowSES$) plus interaction terms with all existing terms that have phase and linguistic group indicators:

\begin{multline}\label{eq3}
 Y_{ijkt} = \alpha + \sum_{m=1}^{2} (\beta_{1m} Phase_{m,t} + \beta_{2m} Phase_{m,t} \ast LingDis_j + \beta_{3m} Phase_{m,t} \ast LowSES_j \\+ \beta_{4m} Phase_{m,t} \ast LingDis_j \ast LowSES_j) + \beta_5 LangDis_j + \beta_6 LowSES_j + \beta_7 LangDis_j \ast LowSES_j \\+\sum_{n=1}^{27}\phi_{n} Topic_{n,i} + \theta_k + \mu_{jt} + \epsilon_{ijkt}
\end{multline}
where the estimated coefficients of various interaction terms captured subtle group differences in the change in writing quality.

\subsection{Additional Methods for Robustness Check}
We performed a series of robustness checks by running Models \ref{eq1}--\ref{eq3} with alternative outcome variables or samples. First, we used each of the seven individual linguistic variables as the regression outcomes, instead of the three composite indices.

Second, we constructed two subsamples of submissions to more ``important'' writing assignments. About 45\% of the assignments gave either 0 or full credit, which may suggest that grading was based on completion rather than quality, so we removed submissions to these assignments from the main analytical sample, keeping writing submissions with greater variability in grades. This gave us 690,554 posts created by 12,905 students across 1,690 courses.

We also calculated the median word count across all submissions to each assignment and found a clear cutoff of 100 words in the distribution, which might be a common length requirement. Therefore, our second subsample only included submissions to assignments with a median length over 100 words, which retained 685,923 posts created by 13,865 students across 1,888 courses.

Lastly, we constructed an alternative sample that included all academic terms through the COVID-19 pandemic. Compared to the main analytical sample, we extended the time span to cover writing submissions from Fall 2019 to Winter 2024 and did not limit to students who had submissions both before and after LLM availability. This sample included 3,365,845 posts created by 59,734 students across 4,416 courses offered over 14 terms. In the regression analyses, we added a binary indicator of pandemic-affected academic terms, which equal to one for all fully online terms between Spring 2020 and Spring 2021, to control for the impact of the pandemic on students' educational experience.

\printbibliography

\end{document}


\maketitle
\begin{table}[ht]
    \centering
    \begin{threeparttable}
    \caption{\textbf{Estimated changes in academic writing proficiency, overall patterns}}\label{tab:s1} 
    \begin{tabular}{lccc}
    \toprule
         & Readability & Lexical Diversity & Syntactic Complexity \\ 
        \midrule
        Phase1 & 0.009& 0.037.& 0.039\\ 
             & (0.021)& (0.021)& (0.026)\\ 
        Phase2 & 0.162***& 0.129***& 0.086*\\ 
             & (0.032)& (0.028)& (0.039)\\
        \midrule
 Assignment content& X& X&X\\
 Course FE& X& X&X\\
 Student FE& X& X&X\\
 Year of study& X& X&X\\ 
\midrule
Adjusted $R^2$ & 0.286& 0.294& 0.318\\ 
 $N_{submission}$& 1,138,337& 1,138,337&1,138,337\\
 $N_{student}$& 16,784& 16,784&16,784\\
   \bottomrule
    \end{tabular}
\begin{tablenotes}
      \footnotesize
      \item\textbf{Notes:} Each column reports results from a separate linear regression at the submission level, where the column header indicates the outcome variable, measured in standard deviations. Phase 1: January 2023 to June 2023; Phase 2: October 2023 to March 2024. Assignment content: All topic weights produced by Latent Dirichlet Allocation for the corresponding assignment. Course FE: Fixed effects for unique course codes (e.g., BIO 101). Student FE: Fixed effects for unique students. Year of study: Fixed effects for the cumulative years of study (in whole numbers) of the corresponding student when creating the submission. Statistical significance is denoted by asterisks: $p<0.10(\cdot)$, $p<0.05(^*)$, $p<0.01(^{**})$, $p<0.001(^{***})$. 
\end{tablenotes}
\end{threeparttable}
\end{table}

\begin{landscape}
\begin{table}[ht]
    \centering    
    \begin{threeparttable}
        \caption{\textbf{Estimated changes in academic writing proficiency, by linguistic background}}\label{tab:s2} 
    \begin{tabular}{lccccccclccc}
    \toprule
    & \multicolumn{3}{c}{Readability} & \multicolumn{1}{c}{} & \multicolumn{3}{c}{Lexical diversity} &  &\multicolumn{3}{c}{Syntactic complexity}\\ 
    \midrule
    ~ & (a)& (b)& (c)& & (a)& (b)& (c)&  &(a)& (b)& (c)\\ 
    \midrule        
       LingDis & -0.111*** & -0.056*** & -0.209*** & & -0.101*** & -0.043*** & -0.151*** &  &-0.177*** & -0.046*** & -0.092***\\ 
       ~ & (0.014) & (0.007) & (0.014) & & (0.010) & (0.005) & (0.010) &  &(0.016) & (0.007) & (0.011)\\ 
       Phase1 & 0.048** & 0.025 & 0.039 & & 0.016 & 0.006 & 0.018 &  &0.019 & 0.012 & 0.018\\ 
       ~ & (0.018) & (0.015) & (0.024) & & (0.016) & (0.017) & (0.021) &  &(0.020) & (0.021) & (0.026)\\ 
       Phase2 & 0.207*** & 0.172*** & 0.194*** & & 0.089*** & 0.070*** & 0.091*** &  &0.063* & 0.051. & 0.050\\ 
       ~ & (0.021) & (0.019) & (0.027) & & (0.020) & (0.020) & (0.028) &  &(0.026) & (0.026) & (0.033)\\ 
       Phase1*LingDis & 0.132*** & 0.040** & 0.027 & & 0.029* & 0.015. & 0.028* &  &0.068*** & 0.016 & 0.040**\\ 
       ~ & (0.020) & (0.012) & (0.023) & & (0.013) & (0.009) & (0.014) &  &(0.018) & (0.010) & (0.015)\\ 
       Phase2*LingDis & 0.298*** & 0.059*** & 0.067* & & 0.137*** & 0.040*** & 0.046* &  &0.094*** & 0.014 & 0.034.\\ 
       ~ & (0.024) & (0.015) & (0.028) & & (0.015) & (0.009) & (0.018) &  &(0.017) & (0.011) & (0.019)\\
    \midrule
       Assignment content & X & X & X & & X & X & X &  &X & X & X\\
       Course FE & X & X & X & & X & X & X &  &X & X & X\\
       Year of study & X & X & X & & X & X & X &  &X & X & X\\ 
    \midrule
       Adjusted $R^2$ & 0.121 & 0.122 & 0.152 & & 0.153 & 0.214 & 0.214 &  &0.253 & 0.254 & 0.231\\ 
       $N_{submission}$ & 1,138,306 & 770,473 & 198,260 & & 1,138,306 & 770,473 & 198,260 &  &1,138,306 & 770,473 & 198,260\\ 
       $N_{student}$ & 16,784 & 11,141 & 2,829 & & 16,784 & 11,141 & 2,829 &  &16,784 & 11,141 & 2,829\\ 
    \bottomrule
    \end{tabular}
\begin{tablenotes}
\footnotesize
    \item \textbf{Notes:} Each column reports results from a separate linear regression at the submission level, where the outer column header indicates the outcome variable, measured in standard deviations. Inner column headers indicate specific linguistically disadvantaged groups (LingDis) analyzed in the corresponding regression: (a) international students, (b) students speaking non-English at home, and (c) students with entering writing score in the 4th quartile. Phase 1: January 2023 to June 2023; Phase 2: October 2023 to March 2024. Assignment content: All topic weights produced by Latent Dirichlet Allocation for the corresponding assignment. Course FE: Fixed effects for unique course codes (e.g., BIO 101). Year of study: Fixed effects for the cumulative years of study (in whole numbers) of the corresponding student when creating the submission. Sample sizes differ across regressions due to different patterns of missingness in student-level attributes. Statistical significance is denoted by asterisks: $p<0.10(\cdot)$, $p<0.05(^*)$, $p<0.01(^{**})$, $p<0.001(^{***})$. 
\end{tablenotes}
\end{threeparttable}
\end{table}
\end{landscape}

\begin{landscape}
\begin{table}[ht]
 \begin{threeparttable}
 \caption {\textbf{Estimated changes in academic writing proficiency, by linguistic background and socioeconomic status (SES)}}\label{tab:s3}
    \small 
    \renewcommand{\arraystretch}{0.9}
    \setlength{\tabcolsep}{3pt}
    \begin{tabular}{lcccccccccccc}
    \toprule
    &\multicolumn{4}{c}{Readability}&\multicolumn{4}{c}{Lexical Diversity}& \multicolumn{4}{c}{Syntactic Complexity}\\ 
    \midrule
        & (a) & (b)) & (c)  &(d)& (a) & (b)) & (c)  &(d)& (a) & (b)) & (c)  &(d)\\ 
    \midrule
 LingDis& -0.086***& -0.035*& -0.210***& -0.264***& -0.066***& -0.039***& -0.167***& -0.159***& -0.088***& -0.045***& -0.109***&-0.139***\\
 & (0.010)& (0.014)& (0.018)& (0.020)& (0.009)& (0.009)& (0.013)& (0.015)& (0.011)& (0.012)& (0.014)&(0.016)\\
 LowSES& -0.046***& -0.059***& -0.008& -0.071***& -0.034***& -0.028***& -0.011& 0.005& -0.026***& -0.012*& 0.009&-0.019\\
 & (0.008)& (0.008)& (0.017)& (0.018)& (0.005)& (0.005)& (0.015)& (0.013)& (0.006)& (0.005)& (0.018)&(0.016)\\
 LingDis*LowSES& 0.063***& -0.018& 0.008& 0.104***& 0.048***& 0.000& 0.031.& 0.007& 0.082***& 0.002& 0.025&0.072***\\
 & (0.014)& (0.019)& (0.023)& (0.026)& (0.012)& (0.010)& (0.019)& (0.018)& (0.014)& (0.013)& (0.022)&(0.021)\\
 Phase1& 0.015& 0.008& 0.028& 0.044& 0.001& 0.000& 0.017& 0.024& 0.001& 0.006& 0.017&0.023\\
 & (0.016)& (0.017)& (0.026)& (0.027)& (0.018)& (0.018)& (0.022)& (0.024)& (0.021)& (0.022)& (0.028)&(0.031)\\
 Phase2& 0.167***& 0.157***& 0.162***& 0.193***& 0.061**& 0.059**& 0.077**& 0.102***& 0.041& 0.040& 0.054&0.063*\\
 & (0.020)& (0.022)& (0.030)& (0.032)& (0.020)& (0.021)& (0.029)& (0.027)& (0.028)& (0.027)& (0.034)&(0.032)\\
 Phase1*LingDis& 0.064***& 0.059*& 0.068*& 0.082**& 0.020& 0.017& 0.049**& 0.033& 0.031*& 0.021& 0.058**&0.067**\\
 & (0.018)& (0.023)& (0.029)& (0.031)& (0.014)& (0.015)& (0.019)& (0.023)& (0.015)& (0.017)& (0.021)&(0.025)\\
 Phase2*LingDis& 0.060*& 0.051.& 0.099**& 0.097.& 0.062***& 0.053***& 0.091***& 0.051.& 0.040*& 0.023& 0.060**&0.057*\\
 & (0.025)& (0.028)& (0.038)& (0.051)& (0.015)& (0.016)& (0.023)& (0.028)& (0.017)& (0.018)& (0.021)&(0.028)\\
 Phase1*LowSES& 0.021.& 0.025*& 0.037& 0.003& 0.011& 0.009& 0.005& -0.014& 0.023*& 0.011& 0.016&-0.001\\
 & (0.011)& (0.012)& (0.040)& (0.039)& (0.008)& (0.009)& (0.029)& (0.022)& (0.010)& (0.009)& (0.028)&(0.026)\\
 Phase2*LowSES& 0.009& 0.021& 0.094*& 0.023& 0.020*& 0.018.& 0.046& -0.024& 0.022.& 0.021*& -0.004&-0.006\\
 & (0.015)& (0.016)& (0.047)& (0.046)& (0.009)& (0.009)& (0.029)& (0.032)& (0.012)& (0.011)& (0.030)&(0.031)\\
 Phase1*LingDis*LowSES& -0.047.& -0.032& -0.085.& -0.073& -0.012& -0.005& -0.037& 0.000& -0.037.& -0.010& -0.040&-0.036\\
 & (0.026)& (0.030)& (0.045)& (0.045)& (0.017)& (0.018)& (0.035)& (0.032)& (0.020)& (0.020)& (0.032)&(0.035)\\
 Phase2*LingDis*LowSES& -0.004& 0.008& -0.098.& -0.050& -0.043.& -0.023& -0.097*& 0.006& -0.056*& -0.019& -0.042&-0.029\\
 & (0.036)& (0.034)& (0.058)& (0.063)& (0.022)& (0.019)& (0.039)& (0.041)& (0.023)& (0.021)& (0.034)&(0.039)\\
\midrule
 Assignment content& X& X& X& X& X& X& X& X& X& X& X&X\\
 Course FE& X& X& X& X& X& X& X& X& X& X& X&X\\
 Year of study& X& X& X& X& X& X& X& X& X& X& X&X\\
\midrule
 Adjusted $R^2$& 0.121& 0.122& 0.152& 0.153& 0.214& 0.214& 0.253& 0.254& 0.231& 0.231& 0.244&0.247\\
 $N_{submission}$ & 762,182& 769,116& 198,072& 193,154& 762,182& 769,116& 198,072& 193,154& 762,182& 769,116& 198,072&193,154\\
 $N_{student}$ & 10,952& 11,111& 2,825& 2,763& 10,952& 11,111& 2,825& 2,763& 10,952& 11,111& 2,825&2,763\\
   \bottomrule
    \end{tabular}
\begin{tablenotes}
    \footnotesize
    \item \textbf{Notes:} Each column reports results from a separate linear regression at the submission level, where the outer column header indicates the outcome variable, measured in standard deviations. Inner column headers indicate specific linguistically disadvantaged (LingDis) and low-SES groups analyzed in the corresponding regression: (a) students speaking non-English at home and students from low-income families, (b) students speaking non-English at home and first-generation college students, (c) students with entering writing score in the 4th quartile and students from low-income families, and (d) students with entering writing score in the 4th quartile, and first-generation college students. Phase 1: January 2023 to June 2023; Phase 2: October 2023 to March 2024. Assignment content: All topic weights produced by Latent Dirichlet Allocation for the corresponding assignment. Course FE: Fixed effects for unique course codes (e.g., BIO 101). Year of study: Fixed effects for the cumulative years of study (in whole numbers) of the corresponding student when creating the submission. Sample sizes differ across regressions due to different patterns of missingness in student-level attributes. Statistical significance is denoted by asterisks: $p<0.10(\cdot)$, $p<0.05(^*)$, $p<0.01(^{**})$, $p<0.001(^{***})$. 
\end{tablenotes}
\end{threeparttable}
\end{table}
\end{landscape}

\begin{table}[ht]
    \centering
    \begin{threeparttable}
    \caption{\textbf{Estimated changes in academic writing proficiency, overall patterns, with individual linguistic measures}}\label{tab:s4} 
    \small 
    \renewcommand{\arraystretch}{0.9}
    \setlength{\tabcolsep}{3pt}
    \begin{tabular}{lccccccc}
    \toprule
    & FK Grade & ARI & Dale-Chall & CTTR & MTLD & Max Depth & Clauses\\
    \midrule
    Phase1 & 0.000& -0.002& 0.030& 0.041& 0.032**& 0.022& 0.056\\ 
         & (0.021)& (0.021)& (0.028)& (0.035)& (0.012)& (0.021)& (0.034)\\
    Phase2 & 0.088**& 0.114***& 0.283***& 0.127**& 0.132***& 0.079**& 0.093.\\ 
         & (0.029)& (0.029)& (0.047)& (0.049)& (0.014)& (0.028)& (0.054)\\
        \midrule
 Assignment content& X& X& X& X& X& X&X\\
 Course FE& X& X& X& X& X& X&X\\
 Student FE& X& X& X& X& X& X&X\\
 Year of study& X& X& X& X& X& X&X\\
  \midrule
    Adjusted $R^2$& 0.271& 0.272& 0.280& 0.414& 0.100& 0.254& 0.286\\ 
 $N_{submission}$& 1,138,337& 1,138,337& 1,138,337& 1,138,337& 1,138,337& 1,138,337&1,138,337\\
 $N_{student}$& 16,784& 16,784& 16,784& 16,784& 16,784& 16,784&16,784\\
     \bottomrule
    \end{tabular}
    \begin{tablenotes}
      \footnotesize
      \item\textbf{Notes:} Each column reports results from a separate linear regression at the submission level, where the column header indicates the outcome variable, measured in standard deviations. FK Grade: Flesch-Kincaid grade level; ARI: Automated Readability Index; Dale-Chall: Dale-Chall readability formula; CTTR: Corrected Type-Token-Ratio; MTLD: Measure of Textual Lexical Diversity; Max Depth: Max depth of the dependency tree; Clauses: Number of clauses. Phase 1: January 2023 to June 2023; Phase 2: October 2023 to March 2024. Assignment content: All topic weights produced by Latent Dirichlet Allocation for the corresponding assignment. Course FE: Fixed effects for unique course codes (e.g., BIO 101). Student FE: Fixed effects for unique students. Year of study: Fixed effects for the cumulative years of study (in whole numbers) of the corresponding student when creating the submission. Statistical significance is denoted by asterisks: $p<0.10(\cdot)$, $p<0.05(^*)$, $p<0.01(^{**})$, $p<0.001(^{***})$. 
\end{tablenotes}
\end{threeparttable}
\end{table}

\begin{landscape}
\begin{table}[ht]
\begin{threeparttable}
    \small
    \setlength{\tabcolsep}{2pt}
     \caption{\textbf{Estimated changes in academic writing proficiency, by linguistic background, with individual linguistic measures}}\label{tab:s5} 
    \begin{tabular}{lcccccccccccc}
    \toprule
    &\multicolumn{3}{c}{FK Grade} &\multicolumn{3}{c}{ARI} & \multicolumn{3}{c}{Dale-Chall} & \multicolumn{3}{c}{CTTR} \\ 
\midrule
        ~ & (a) & (b) & (c) & (a) & (b) & (c) &(a) & (b) & (c)  &(a) & (b) & (c)  \\ 
\midrule        
LingDis & -0.173***& -0.063***& -0.189***& -0.161***& -0.057***& -0.194***& 0.001& -0.048***& -0.244***& -0.120***& -0.041***& -0.179***\\
& (0.016)& (0.008)& (0.015)& (0.017)& (0.008)& (0.015)& (0.013)& (0.006)& (0.015)& (0.016)& (0.007)& (0.013)\\
Phase1 & 0.034& 0.008& 0.028& 0.035& 0.012& 0.030& 0.073***& 0.055**& 0.058*& 0.019& 0.011& 0.016\\
& (0.021)& (0.016)& (0.028)& (0.022)& (0.017)& (0.028)& (0.019)& (0.020)& (0.029)& (0.026)& (0.027)& (0.033)\\
Phase2 & 0.140***& 0.108***& 0.135***& 0.167***& 0.134***& 0.164***& 0.315***& 0.274***& 0.282***& 0.092**& 0.077*& 0.088*\\
& (0.020)& (0.019)& (0.032)& (0.021)& (0.019)& (0.030)& (0.032)& (0.031)& (0.047)& (0.032)& (0.032)& (0.039)\\
Phase1*LingDis & 0.136***& 0.037*& 0.024& 0.125***& 0.034*& 0.023& 0.134***& 0.048***& 0.035& 0.038.& 0.008& 0.053**\\
& (0.021)& (0.015)& (0.025)& (0.021)& (0.015)& (0.025)& (0.022)& (0.011)& (0.025)& (0.020)& (0.012)& (0.018)\\
Phase2*LingDis & 0.243***& 0.051***& 0.048& 0.238***& 0.047**& 0.039& 0.413***& 0.078***& 0.114***& 0.116***& 0.025*& 0.061**\\
& (0.022)& (0.015)& (0.030)& (0.021)& (0.015)& (0.030)& (0.035)& (0.019)& (0.030)& (0.021)& (0.011)& (0.023)\\
       \midrule
 Assignment content& X& X& X& X& X& X& X& X& X& X& X&X\\
 Course FE& X& X& X& X& X& X& X& X& X& X& X&X\\
 Year of study& X& X& X& X& X& X& X& X& X& X& X&X\\
        \midrule
Adjusted $R^2$& 0.102& 0.103& 0.129& 0.108& 0.110& 0.139& 0.147& 0.153& 0.185& 0.333& 0.340& 0.360\\
$N_{submission}$& 1,138,306 & 770,473 & 198,260 & 1,138,306 & 770,473 & 198,260 & 1,138,306 & 770,473 & 198,260 & 1,138,306 & 770,473 & 198,260 \\
$N_{student}$& 16,784 & 11,141 & 2,829 & 16,784 & 11,141 & 2,829 & 16,784 & 11,141 & 2,829 & 16,784 & 11,141 & 2,829 \\
\bottomrule
    \end{tabular}
\end{threeparttable}
\end{table}

\begin{table}[ht]
    \raggedright
    \small
    \setlength{\tabcolsep}{2pt}
    \begin{tabular}{llllllllll}
    \toprule
    &\multicolumn{3}{c}{MTLD} &\multicolumn{3}{c}{Max Depth} & \multicolumn{3}{c}{Clause} \\ 
\midrule
        ~ & (a) & (b) & (c) & (a) & (b) & (c) & (a) & (b) & (c)  \\ 
\midrule        
LingDis & -0.082***& -0.045***& -0.123***& -0.248***& -0.067***& -0.116***& -0.107***& -0.025**& -0.068***\\
& (0.008)& (0.005)& (0.009)& (0.015)& (0.008)& (0.013)& (0.018)& (0.008)& (0.011)\\
Phase1 & 0.013& 0.001& 0.020& 0.014& 0.011& 0.034& 0.023& 0.013& 0.003\\
& (0.009)& (0.010)& (0.017)& (0.019)& (0.020)& (0.029)& (0.024)& (0.023)& (0.027)\\
Phase2 & 0.086***& 0.063***& 0.095***& 0.076***& 0.063**& 0.074*& 0.050& 0.040& 0.027\\
& (0.011)& (0.012)& (0.021)& (0.022)& (0.023)& (0.030)& (0.033)& (0.033)& (0.040)\\
Phase1*LingDis & 0.020*& 0.022**& 0.003& 0.086***& 0.019& 0.034.& 0.051**& 0.012& 0.045**\\
& (0.010)& (0.008)& (0.015)& (0.019)& (0.012)& (0.018)& (0.019)& (0.011)& (0.016)\\
Phase2*LingDis & 0.159***& 0.056***& 0.032.& 0.131***& 0.015& 0.032& 0.057**& 0.012& 0.037.\\
& (0.015)& (0.009)& (0.018)& (0.019)& (0.014)& (0.023)& (0.021)& (0.012)& (0.020)\\
       \midrule
 Assignment content& X& X& X& X& X& X& X& X& X\\
 Course FE& X& X& X& X& X& X& X& X& X\\
 Year of study& X& X& X& X& X& X& X& X& X\\
        \midrule
Adjusted $R^2$& 0.057& 0.054& 0.073& 0.158& 0.150& 0.149& 0.225& 0.227& 0.251\\
$N_{submission}$& 1,138,306 & 770,473 & 198,260 & 1,138,306 & 770,473 & 198,260 & 1,138,306 & 770,473 & 198,260 \\
$N_{student}$& 16,784 & 11,141 & 2,829 & 16,784 & 11,141 & 2,829 & 16,784 & 11,141 & 2,829 \\
\bottomrule
    \end{tabular}
   \begin{tablenotes}
\footnotesize
    \item \textbf{Notes:} Each column reports results from a separate linear regression at the submission level, where the outer column header indicates the outcome variable, measured in standard deviations. FK Grade: Flesch-Kincaid grade level; ARI: Automated Readability Index; Dale-Chall: Dale-Chall readability formula; CTTR: Corrected Type-Token-Ratio; MTLD: Measure of Textual Lexical Diversity; Max Depth: Max depth of the dependency tree; Clauses: Number of clauses. Inner column headers indicate specific linguistically disadvantaged groups (LingDis) analyzed in the corresponding regression: (a) international students, (b) students speaking non-English at home, and (c) students with entering writing score in the 4th quartile. Phase 1: January 2023 to June 2023; Phase 2: October 2023 to March 2024. Assignment content: All topic weights produced by Latent Dirichlet Allocation for the corresponding assignment. Course FE: Fixed effects for unique course codes (e.g., BIO 101). Year of study: Fixed effects for the cumulative years of study (in whole numbers) of the corresponding student when creating the submission. Sample sizes differ across regressions due to different patterns of missingness in student-level attributes. Statistical significance is denoted by asterisks: $p<0.10(\cdot)$, $p<0.05(^*)$, $p<0.01(^{**})$, $p<0.001(^{***})$. 
\end{tablenotes}
\label{Table S5} 
\end{table}
\end{landscape}

\begin{landscape}
\begin{table}[ht]
    \small
    \renewcommand{\arraystretch}{0.9}
    \setlength{\tabcolsep}{2pt}
    \caption {\textbf{Estimated changes in academic writing proficiency, by linguistic background and socioeconomic status, with individual linguistic measures}} \label{tab:s6} 
\begin{tabular}{lcccccccccccc}
    \toprule
 & \multicolumn{4}{c}{FK Grade}& \multicolumn{4}{c}{ARI}& \multicolumn{4}{c}{Dale-Chall}\\
     \midrule
& (a)& (b)& (c)& (d)& (a)& (b)& (c)& (d)& (a)& (b)& (c)& (d)\\
     \midrule
LingDis& -0.105***& -0.041*& -0.197***& -0.264***& -0.096***& -0.033.& -0.201***& -0.268***& -0.059***& -0.031**& -0.233***& -0.259***\\
 & (0.012)& (0.018)& (0.020)& (0.022)& (0.011)& (0.018)& (0.020)& (0.022)& (0.010)& (0.012)& (0.020)& (0.021)\\
LowSES& -0.043***& -0.050***& -0.007& -0.077***& -0.041***& -0.047***& -0.007& -0.073**& -0.054***& -0.079***& -0.009& -0.061***\\
 & (0.008)& (0.009)& (0.020)& (0.022)& (0.008)& (0.009)& (0.020)& (0.022)& (0.009)& (0.009)& (0.018)& (0.017)\\
LingDis*LowSES& 0.083***& -0.021& 0.018& 0.133***& 0.076***& -0.026& 0.017& 0.130***& 0.030*& -0.006& -0.013& 0.050.\\
 & (0.015)& (0.023)& (0.027)& (0.031)& (0.015)& (0.024)& (0.026)& (0.030)& (0.014)& (0.015)& (0.025)& (0.026)\\
Phase1& -0.002& -0.008& 0.014& 0.029& 0.002& -0.004& 0.016& 0.035& 0.046*& 0.036.& 0.054.& 0.067*\\
 & (0.017)& (0.018)& (0.031)& (0.031)& (0.017)& (0.018)& (0.030)& (0.030)& (0.021)& (0.021)& (0.031)& (0.033)\\
Phase2& 0.103***& 0.094***& 0.112**& 0.136***& 0.129***& 0.121***& 0.138***& 0.169***& 0.269***& 0.254***& 0.236***& 0.273***\\
 & (0.021)& (0.023)& (0.035)& (0.038)& (0.020)& (0.022)& (0.033)& (0.035)& (0.030)& (0.031)& (0.049)& (0.050)\\
Phase1*LingDis& 0.062**& 0.050.& 0.069*& 0.092**& 0.056**& 0.048.& 0.070*& 0.087*& 0.074***& 0.080***& 0.064*& 0.068*\\
 & (0.021)& (0.028)& (0.030)& (0.035)& (0.020)& (0.028)& (0.030)& (0.034)& (0.017)& (0.023)& (0.033)& (0.033)\\
Phase2*LingDis& 0.053*& 0.029& 0.066.& 0.082& 0.046.& 0.029& 0.062& 0.072& 0.082*& 0.095**& 0.169***& 0.135*\\
 & (0.025)& (0.031)& (0.040)& (0.054)& (0.024)& (0.031)& (0.039)& (0.052)& (0.035)& (0.031)& (0.047)& (0.057)\\
Phase1*LowSES& 0.022.& 0.025.& 0.051& 0.017& 0.020& 0.024.& 0.054& 0.007& 0.021.& 0.027*& 0.006& -0.014\\
 & (0.013)& (0.013)& (0.047)& (0.046)& (0.012)& (0.013)& (0.047)& (0.045)& (0.012)& (0.013)& (0.038)& (0.034)\\
Phase2*LowSES& 0.009& 0.018& 0.071& 0.024& 0.009& 0.018& 0.082.& 0.017& 0.010& 0.026& 0.129*& 0.029\\
 & (0.015)& (0.016)& (0.048)& (0.050)& (0.014)& (0.016)& (0.048)& (0.052)& (0.020)& (0.021)& (0.057)& (0.052)\\
 Phase1*LingDis*LowSES& -0.048& -0.022& -0.101.& -0.095.& -0.044& -0.024& -0.106*& -0.087& -0.050*& -0.051.& -0.048& -0.036\\
 & (0.029)& (0.035)& (0.052)& (0.054)& (0.029)& (0.034)& (0.051)& (0.053)& (0.025)& (0.030)& (0.048)& (0.042)\\
 Phase2*LingDis*LowSES& -0.005& 0.031& -0.064& -0.056& -0.001& 0.025& -0.076& -0.051& -0.008& -0.031& -0.154.& -0.042\\
 & (0.035)& (0.037)& (0.058)& (0.069)& (0.035)& (0.037)& (0.058)& (0.069)& (0.045)& (0.037)& (0.079)& (0.068)\\
      \midrule
 Assignment content& X& X& X& X& X& X& X& X& X& X& X&X\\
 Course FE& X& X& X& X& X& X& X& X& X& X& X&X\\
 Year of study& X& X& X& X& X& X& X& X& X& X& X&X\\
      \midrule
 Adjusted $R^2$& 0.104& 0.104& 0.129& 0.130& 0.111& 0.111& 0.139& 0.140& 0.154& 0.155& 0.185& 0.186\\
 $N_{submission}$& 762,182& 769,116& 198,072& 193,154& 762,182& 769,116& 198,072& 193,154& 762,182& 769,116& 198,072& 193,154\\
 $N_{student}$& 10,952& 11,111& 2,825& 2,763& 10,952& 11,111& 2,825& 2,763& 10,952& 11,111& 2,825& 2,763\\
   \bottomrule
\end{tabular}
\end{table}

\begin{table}[ht]
    \raggedright
    \small
    \renewcommand{\arraystretch}{0.9}
    \setlength{\tabcolsep}{2pt}
\begin{tabular}{lllllllllllll}
    \toprule
     &\multicolumn{4}{c}{CTTR}& \multicolumn{4}{c}{MTLD}& \multicolumn{4}{c}{Max Depth}\\
    \midrule
    & (a)& (b)& (c)& (d)& (a)& (b)& (c)& (d)& (a)& (b)& (c)& (d)\\
    \midrule
    LingDis&-0.069***&-0.031**&-0.196***&-0.198***& -0.062***& -0.046***& -0.138***& -0.120***& -0.127***& -0.068***& -0.140***& -0.181***\\
 &(0.012)&(0.011)&(0.017)&(0.020)& (0.008)& (0.009)& (0.012)& (0.013)& (0.012)& (0.014)& (0.016)& (0.018)\\
    LowSES&-0.048***&-0.035***&-0.003&-0.003& -0.020***& -0.021***& -0.019& 0.013& -0.029***& -0.014*& 0.004& -0.031\\
     &(0.008)&(0.007)&(0.021)&(0.017)& (0.004)& (0.004)& (0.013)& (0.012)& (0.007)& (0.007)& (0.022)& (0.020)\\
    LingDis*LowSES&0.061***&-0.007&0.029&0.026& 0.035***& 0.007& 0.034.& -0.011& 0.113***& 0.005& 0.039& 0.101***\\
     &(0.017)&(0.013)&(0.026)&(0.024)& (0.009)& (0.010)& (0.017)& (0.016)& (0.016)& (0.016)& (0.026)& (0.024)\\
    Phase1&0.001&-0.002&0.012&0.014& 0.002& 0.001& 0.023& 0.034& 0.000& 0.003& 0.034& 0.053\\
     &(0.028)&(0.028)&(0.035)&(0.037)& (0.011)& (0.012)& (0.021)& (0.021)& (0.021)& (0.022)& (0.031)& (0.033)\\
    Phase2&0.061.&0.060.&0.081*&0.102**& 0.061***& 0.057***& 0.073**& 0.103***& 0.051*& 0.048*& 0.075*& 0.099**\\
     &(0.033)&(0.033)&(0.041)&(0.038)& (0.013)& (0.013)& (0.022)& (0.022)& (0.023)& (0.024)& (0.031)& (0.033)\\
    Phase1*LingDis&0.015&0.011&0.074**&0.068*& 0.026.& 0.023& 0.025& -0.001& 0.041*& 0.030& 0.063*& 0.058.\\
     &(0.017)&(0.018)&(0.024)&(0.030)& (0.014)& (0.016)& (0.020)& (0.022)& (0.019)& (0.022)& (0.026)& (0.030)\\
    Phase2*LingDis&0.054**&0.036.&0.106***&0.058& 0.069***& 0.069***& 0.076**& 0.044& 0.045*& 0.022& 0.059*& 0.059\\
     &(0.020)&(0.021)&(0.029)&(0.036)& (0.016)& (0.015)& (0.023)& (0.028)& (0.020)& (0.022)& (0.028)& (0.036)\\
    Phase1*LowSES&0.021.&0.020*&0.022&0.005& 0.000& -0.003& -0.012& -0.032& 0.024.& 0.014& 0.010& -0.029\\
     &(0.012)&(0.010)&(0.034)&(0.028)& (0.008)& (0.011)& (0.035)& (0.025)& (0.013)& (0.012)& (0.035)& (0.032)\\
    Phase2*LowSES&0.036**&0.028*&0.025&-0.024& 0.004& 0.008& 0.068*& -0.024& 0.027*& 0.028*& 0.005& -0.029\\
     &(0.013)&(0.013)&(0.036)&(0.039)& (0.009)& (0.009)& (0.031)& (0.034)& (0.012)& (0.012)& (0.039)& (0.036)\\
     Phase1*LingDis*LowSES&-0.019&-0.008&-0.046&-0.021& -0.006& -0.001& -0.028& 0.022& -0.050*& -0.019& -0.053& -0.019\\
     &(0.024)&(0.022)&(0.041)&(0.042)& (0.017)& (0.017)& (0.038)& (0.030)& (0.025)& (0.026)& (0.041)& (0.041)\\
     Phase2*LingDis*LowSES&-0.064*&-0.025&-0.087.&0.015& -0.021& -0.021& -0.107**& -0.003& -0.065*& -0.017& -0.048& -0.028\\
     &(0.029)&(0.026)&(0.047)&(0.050)& (0.021)& (0.018)& (0.041)& (0.042)& (0.026)& (0.025)& (0.044)& (0.049)\\
         \midrule
 Assignment content& X& X& X& X& X& X& X& X& X& X& X&X\\
 Course FE& X& X& X& X& X& X& X& X& X& X& X&X\\
 Year of study& X& X& X& X& X& X& X& X& X& X& X&X\\
     \midrule
     Adjusted $R^2$&0.104&0.104&0.129&0.130& 0.111& 0.111& 0.139& 0.140& 0.154& 0.155& 0.185& 0.186\\
     $N_{submission}$&762,182&769,116&198,072&193,154& 762,182& 769,116& 198,072& 193,154& 762,182& 769,116& 198,072& 193,154\\
     $N_{student}$&10,952&11,111&2,825&2,763& 10,952& 11,111& 2,825& 2,763& 10,952& 11,111& 2,825& 2,763\\
    \bottomrule
\end{tabular}
\end{table}

\begin{table}[ht]
    \raggedright
    \small
    \renewcommand{\arraystretch}{0.9}
    \setlength{\tabcolsep}{2pt}
\begin{tabular}{lllll}
    \toprule
     & \multicolumn{4}{c}{Clause}\\
    \midrule
    & (a)& (b)& (c)& (d)\\
    \midrule
    LingDis& -0.050***& -0.022.& -0.079***& -0.098***\\
 & (0.013)& (0.013)& (0.014)& (0.018)\\
    LowSES& -0.024***& -0.009& 0.014& -0.007\\
     & (0.006)& (0.006)& (0.017)& (0.017)\\
    LingDis*LowSES& 0.052***& -0.002& 0.012& 0.044.\\
     & (0.016)& (0.014)& (0.023)& (0.024)\\
    Phase1& 0.003& 0.008& -0.001& -0.007\\
     & (0.024)& (0.024)& (0.029)& (0.033)\\
    Phase2& 0.030& 0.032& 0.032& 0.026\\
     & (0.035)& (0.033)& (0.043)& (0.037)\\
    Phase1*LingDis& 0.021& 0.011& 0.053*& 0.076**\\
     & (0.016)& (0.017)& (0.021)& (0.028)\\
    Phase2*LingDis& 0.036.& 0.024& 0.062**& 0.055.\\
     & (0.019)& (0.020)& (0.023)& (0.030)\\
    Phase1*LowSES& 0.021*& 0.008& 0.022& 0.027\\
     & (0.011)& (0.009)& (0.030)& (0.029)\\
    Phase2*LowSES& 0.018& 0.014& -0.012& 0.017\\
     & (0.014)& (0.011)& (0.031)& (0.035)\\
     Phase1*LingDis*LowSES& -0.025& -0.001& -0.026& -0.052\\
     & (0.022)& (0.021)& (0.034)& (0.038)\\
     Phase2*LingDis*LowSES& -0.046.& -0.021& -0.036& -0.031\\
     & (0.026)& (0.026)& (0.037)& (0.041)\\
         \midrule
 Assignment content& X& X& X&X\\
 Course FE& X& X& X&X\\
 Year of study& X& X& X&X\\
     \midrule
     Adjusted $R^2$& 0.227& 0.227& 0.252& 0.253\\
     $N_{submission}$& 762,182& 769,116& 198,072& 193,154\\
     $N_{student}$& 10,952& 11,111& 2,825& 2,763\\
    \bottomrule
\end{tabular}
\begin{tablenotes}
    \footnotesize
    \item \textbf{Notes:} Each column reports results from a separate linear regression at the submission level, where the outer column header indicates the outcome variable, measured in standard deviations. FK Grade: Flesch-Kincaid grade level; ARI: Automated Readability Index; Dale-Chall: Dale-Chall readability formula; CTTR: Corrected Type-Token-Ratio; MTLD: Measure of Textual Lexical Diversity; Max Depth: Max depth of the dependency tree; Clauses: Number of clauses. Inner column headers indicate specific linguistically disadvantaged (LingDis) and low-SES groups analyzed in the corresponding regression: (a) students speaking non-English at home and students from low-income families, (b) students speaking non-English at home and first-generation college students, (c) students with entering writing score in the 4th quartile and students from low-income families, and (d) students with entering writing score in the 4th quartile, and first-generation college students. Phase 1: January 2023 to June 2023; Phase 2: October 2023 to March 2024. Assignment content: All topic weights produced by Latent Dirichlet Allocation for the corresponding assignment. Course FE: Fixed effects for unique course codes (e.g., BIO 101). Year of study: Fixed effects for the cumulative years of study (in whole numbers) of the corresponding student when creating the submission. Sample sizes differ across regressions due to different patterns of missingness in student-level attributes. Statistical significance is denoted by asterisks: $p<0.10(\cdot)$, $p<0.05(^*)$, $p<0.01(^{**})$, $p<0.001(^{***})$. 
\end{tablenotes}
\end{table}
\end{landscape}

\begin{figure}[ht]
\captionsetup[subfigure]{justification=centering}
\centering
\begin{subfigure}{\textwidth}
\centering
\includegraphics[width=0.8\textwidth]{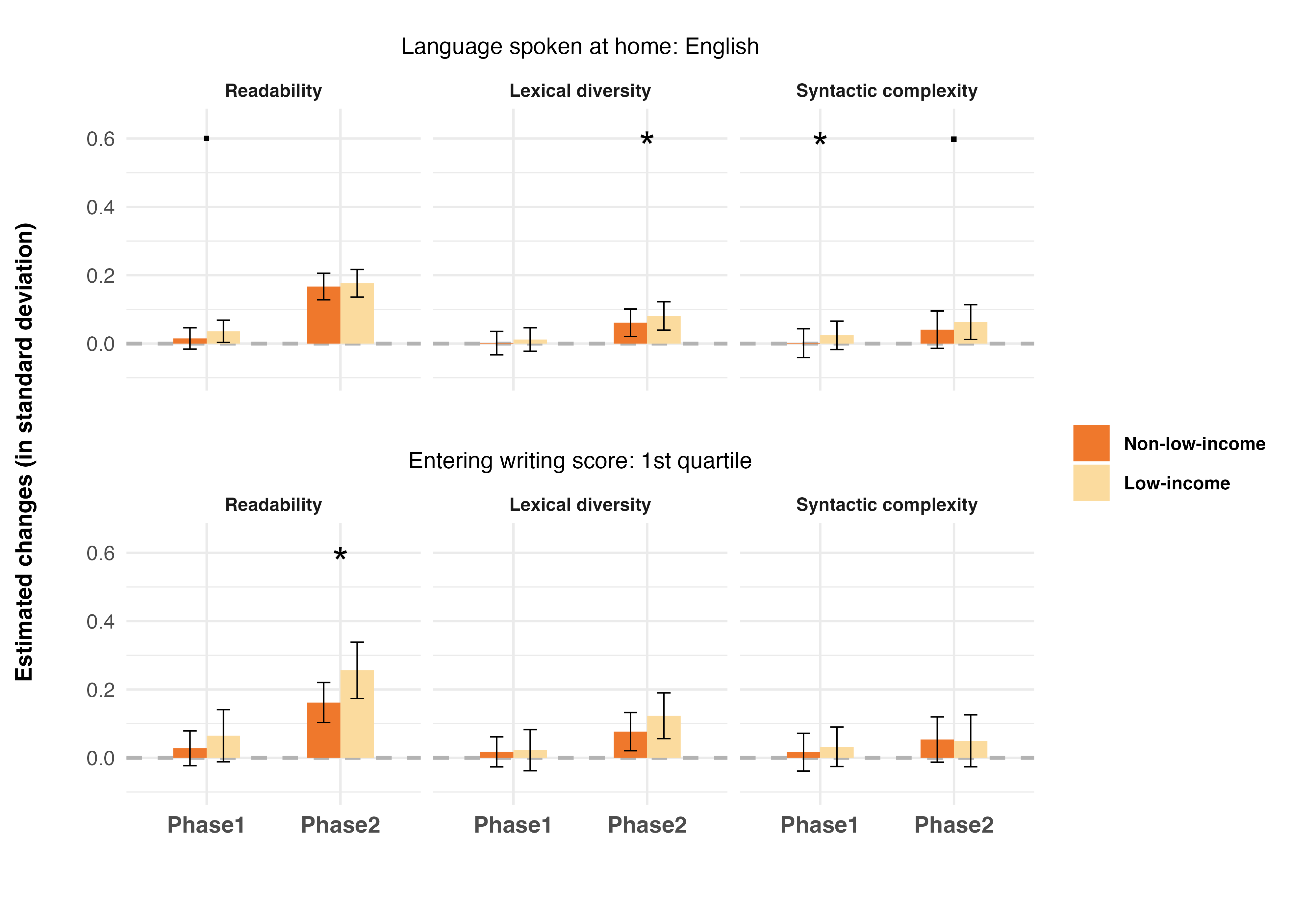}
\caption{SES: Family income}
\label{fig:s1a}
\end{subfigure}
\begin{subfigure}{\textwidth}
\centering
\includegraphics[width=0.8\textwidth]{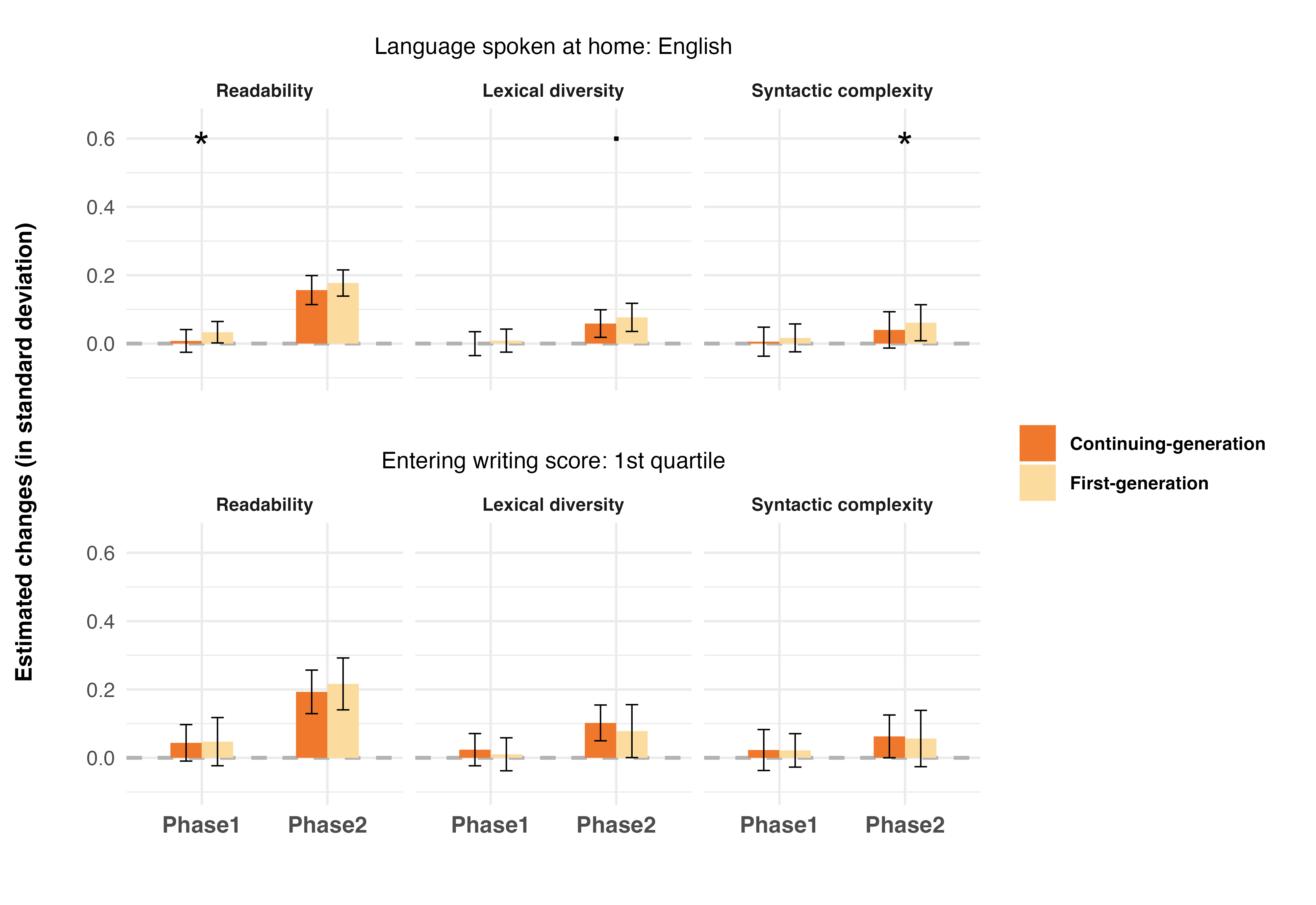}
\centering\caption{SES: Parental education}
\label{fig:s1b}
\end{subfigure}
\caption{
\small
\textbf{Estimated changes in academic writing proficiency among linguistically advantaged student groups, by socioeconomic status (SES).} Each bar shows the predicted average change (in standard deviation) in a composite writing proficiency index between a post-LLM phase and the pre-LLM period for a given student subgroup. Phase 1: January 2023 to June 2023; Phase 2: October 2023 to March 2024. Error bars indicate 90\% confidence intervals. Statistical significance for the difference between adjacent bars (three-way interaction term between indicators of the linguistic group, socioeconomic group and phase in the regression model) is denoted by asterisks: $p<0.10(\cdot)$, $p<0.05(^*)$, $p<0.01(^{**})$, $p<0.001(^{***})$.}
\label{fig:s1}
\end{figure}

\begin{figure}[ht]
    \centering
    \includegraphics[width=0.6\textwidth]{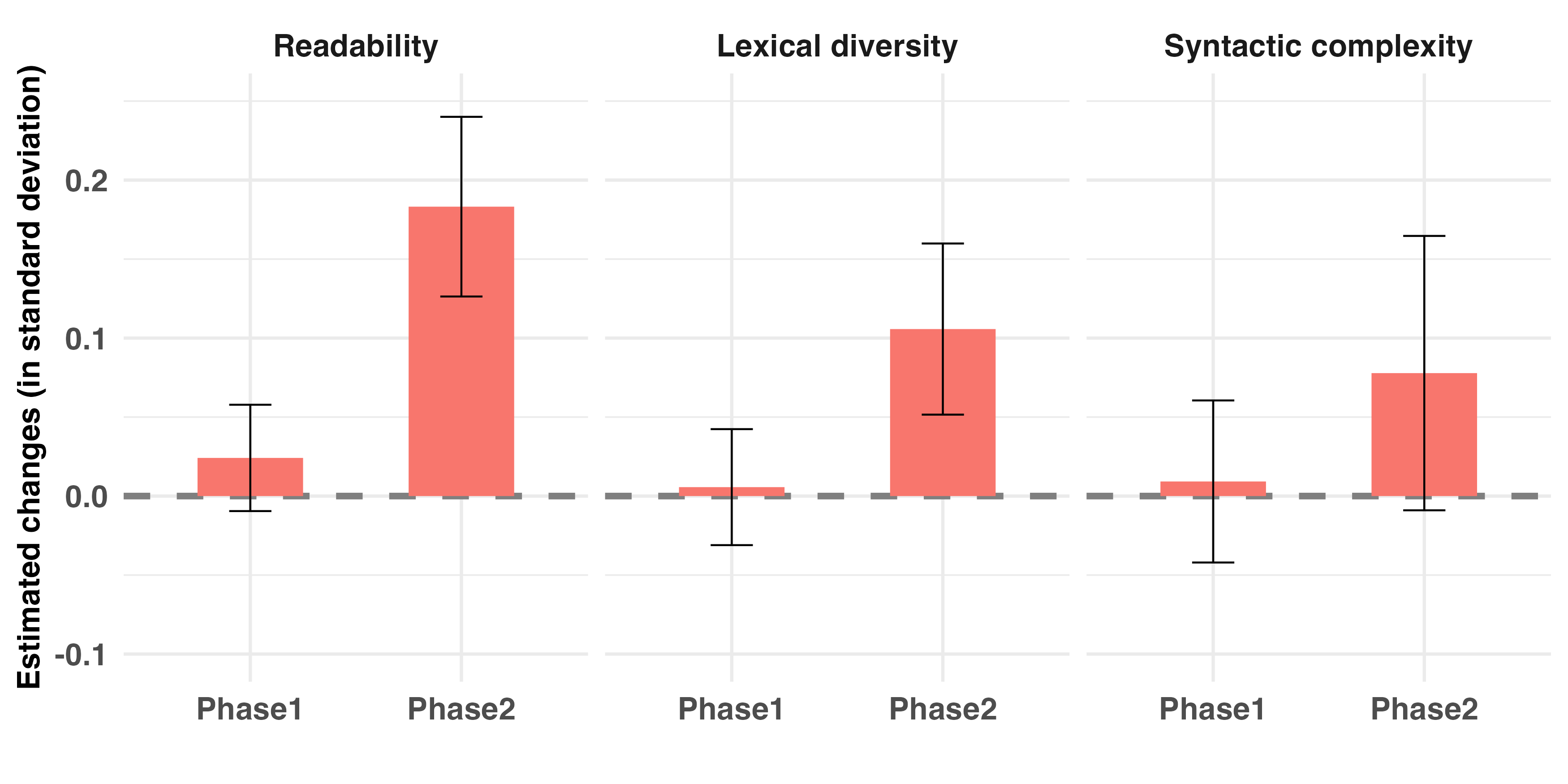}
    \caption{\textbf{Estimated changes in academic writing proficiency, for writing submissions with greater grading variability.} The sample only includes writing submissions to assignments where the grading was not limited to a binary scale such as 0 or full credit. Each bar represents the estimate of the average change (in standard deviation) in a composite writing proficiency index between a post-LLM phase and the pre-LLM period in the data. Phase 1: January 2023 to June 2023; Phase 2: October 2023 to March 2024. Error bars indicate 90\% confidence intervals.}
\label{fig:s2}
\end{figure}

\begin{figure}[ht]
    \centering
    \includegraphics[width=0.85\textwidth]{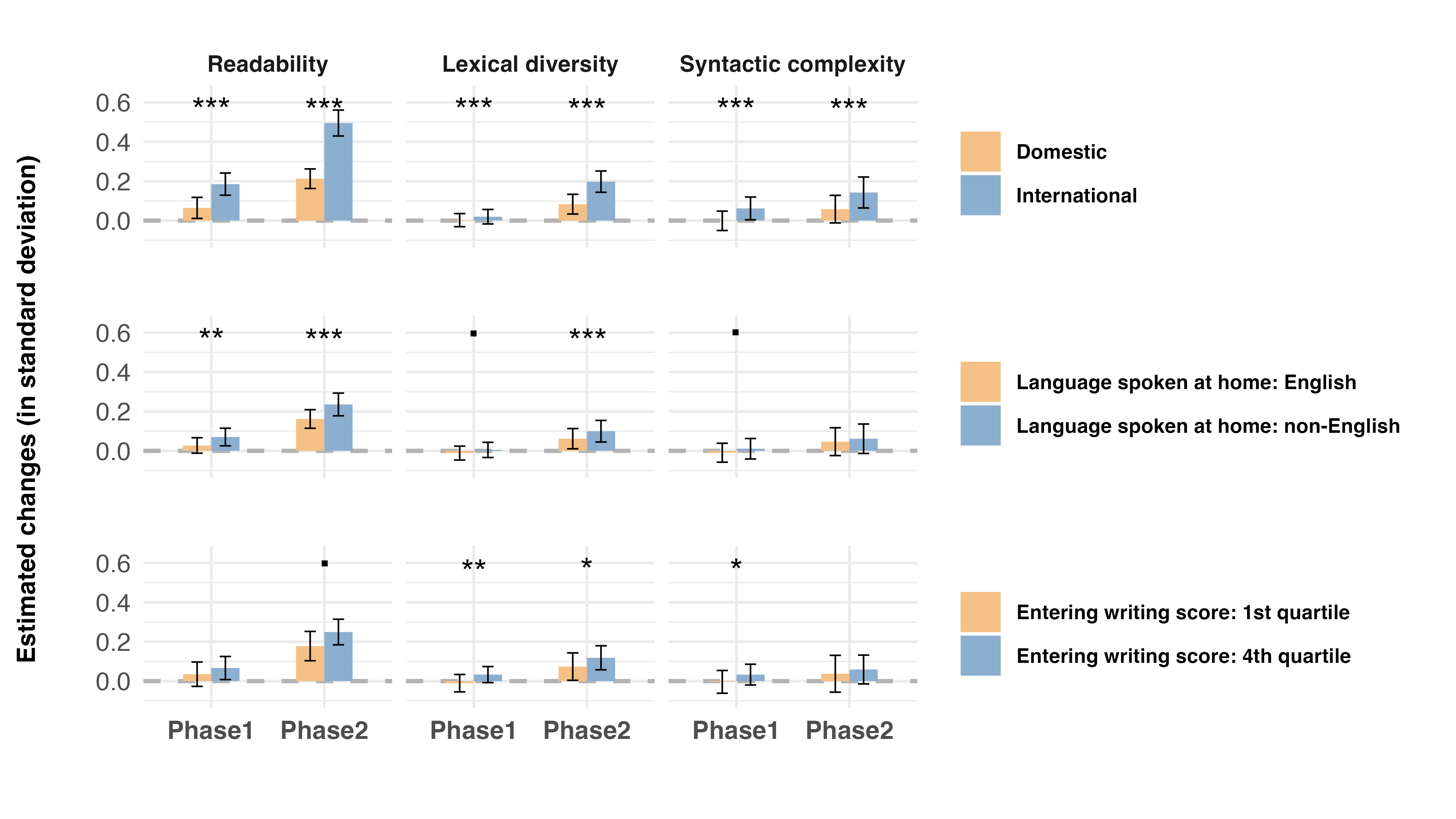}
    \caption{\textbf{Estimated changes in academic writing proficiency, by linguistic background, for writing submissions with greater grading variability.} The sample only includes writing submissions to assignments where the grading was not limited to a binary scale such as 0 or full credit. Each bar shows the predicted average change (in standard deviation) in a composite writing proficiency index between a post-LLM phase and the pre-LLM period for a given student group. Phase 1: January 2023 to June 2023; Phase 2: October 2023 to March 2024. Error bars indicate 90\% confidence intervals. Statistical significance for the difference between adjacent bars (interaction term between the group and the phase indicators in the regression model) is denoted by asterisks: $p<0.10(\cdot)$, $p<0.05(^*)$, $p<0.01(^{**})$, $p<0.001(^{***})$.}
\label{fig:s3}
\end{figure}

\begin{figure}[ht]
\captionsetup[subfigure]{justification=centering}
\centering
\begin{subfigure}{0.85\textwidth}
\centering
\includegraphics[width=0.85\textwidth]{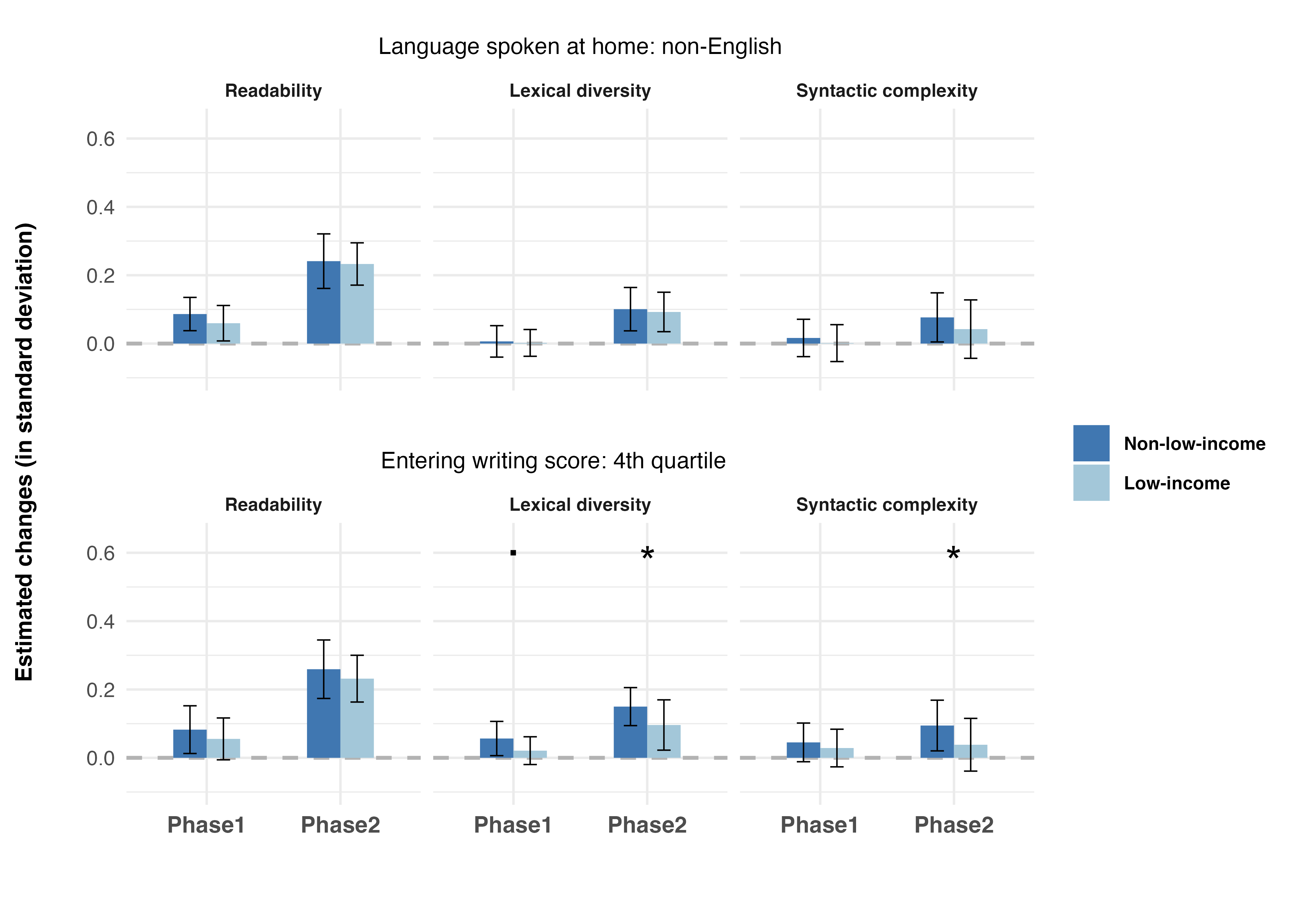}
\caption{SES: Family income}
\label{fig:s4a}
\end{subfigure}
\begin{subfigure}{0.85\textwidth}
\centering
\includegraphics[width=0.85\textwidth]{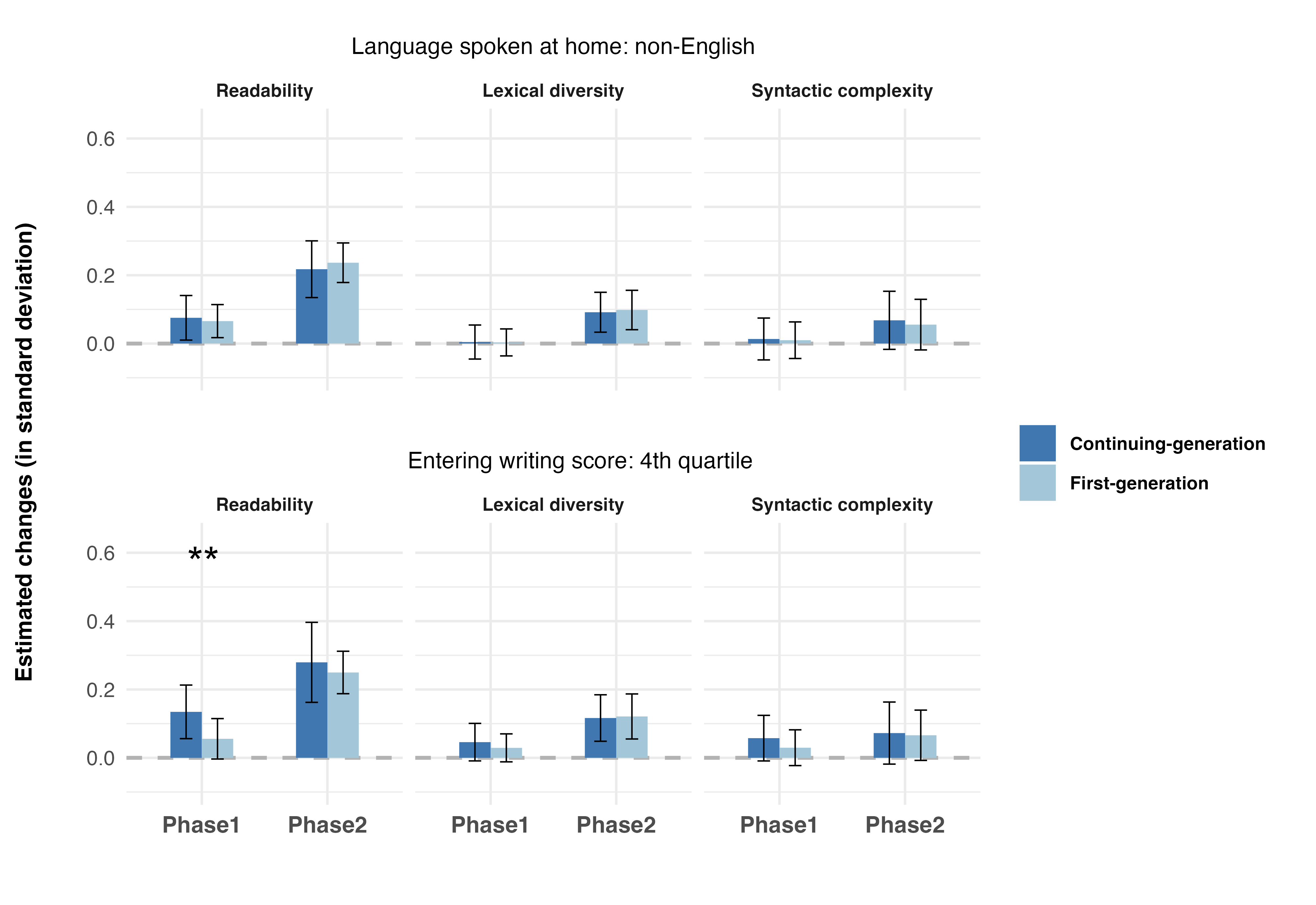}
\centering\caption{SES: Parental education}
\label{fig:s4b}
\end{subfigure}
\caption{
\small
\textbf{Estimated changes in academic writing proficiency among linguistically disadvantaged student groups, by socioeconomic status (SES), for writing submissions with greater grading variability.} The sample only includes writing submissions to assignments where the grading was not limited to a binary scale such as 0 or full credit. Each bar shows the predicted average change (in standard deviation) in a composite writing proficiency index between a post-LLM phase and the pre-LLM period for a given student subgroup. Phase 1: January 2023 to June 2023; Phase 2: October 2023 to March 2024. Error bars indicate 90\% confidence intervals. Statistical significance for the difference between adjacent bars (three-way interaction term between indicators of the linguistic group, socioeconomic group and phase in the regression model) is denoted by asterisks: $p<0.10(\cdot)$, $p<0.05(^*)$, $p<0.01(^{**})$, $p<0.001(^{***})$.}
\label{fig:s4}
\end{figure}

\begin{figure}[ht]
    \centering
    \includegraphics[width=0.6\textwidth]{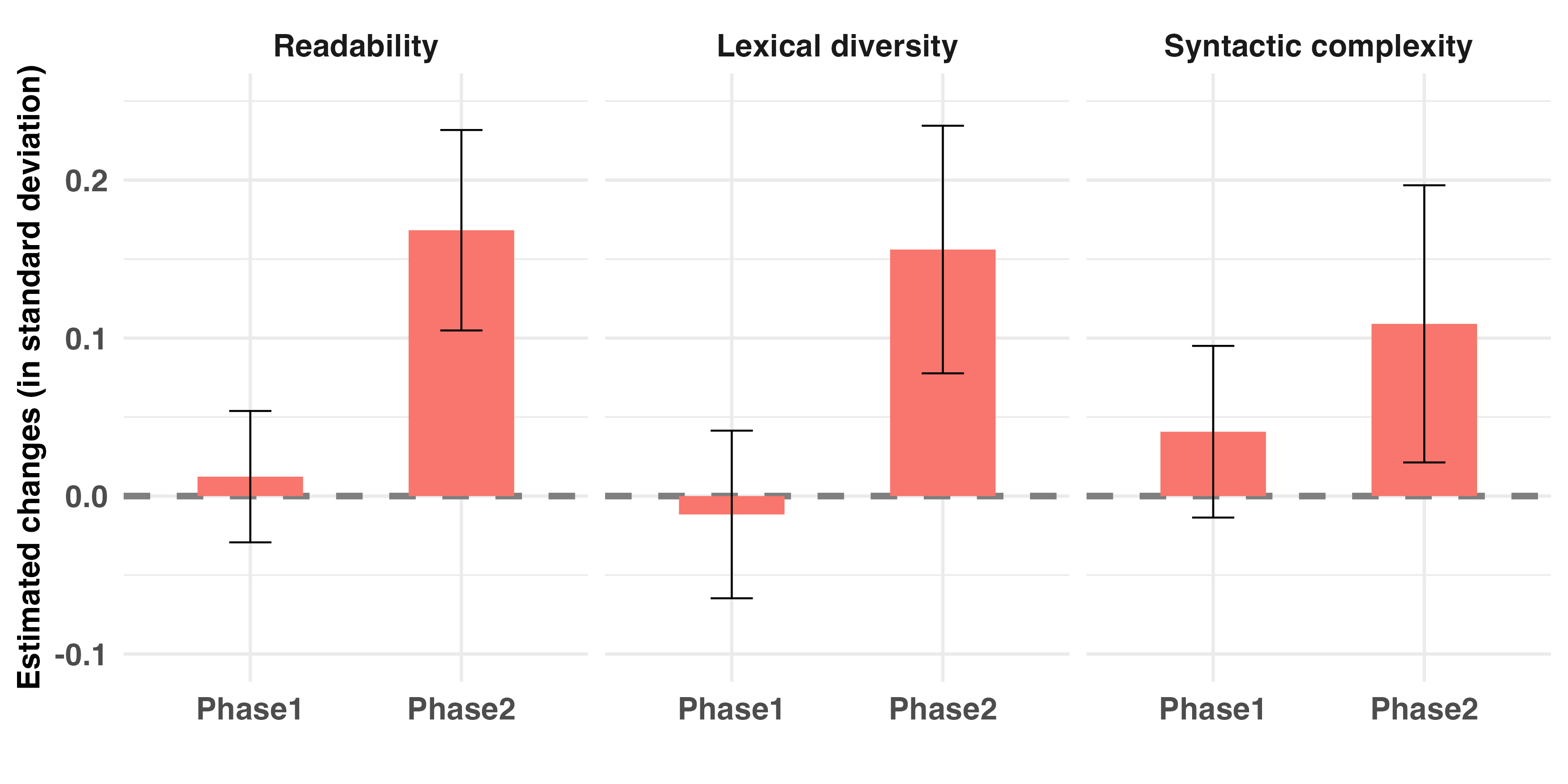}
    \caption{\textbf{Estimated changes in academic writing proficiency, for writing submissions with longer content.} The sample only includes writing submissions to assignments where the median submission length was over 100 words. Each bar represents the estimate of the average change (in standard deviation) in a composite writing proficiency index between a post-LLM phase and the pre-LLM period in the data. Phase 1: January 2023 to June 2023; Phase 2: October 2023 to March 2024. Error bars indicate 90\% confidence intervals.}
\label{fig:s5}
\end{figure}

\begin{figure}[ht]
    \centering
    \includegraphics[width=0.85\textwidth]{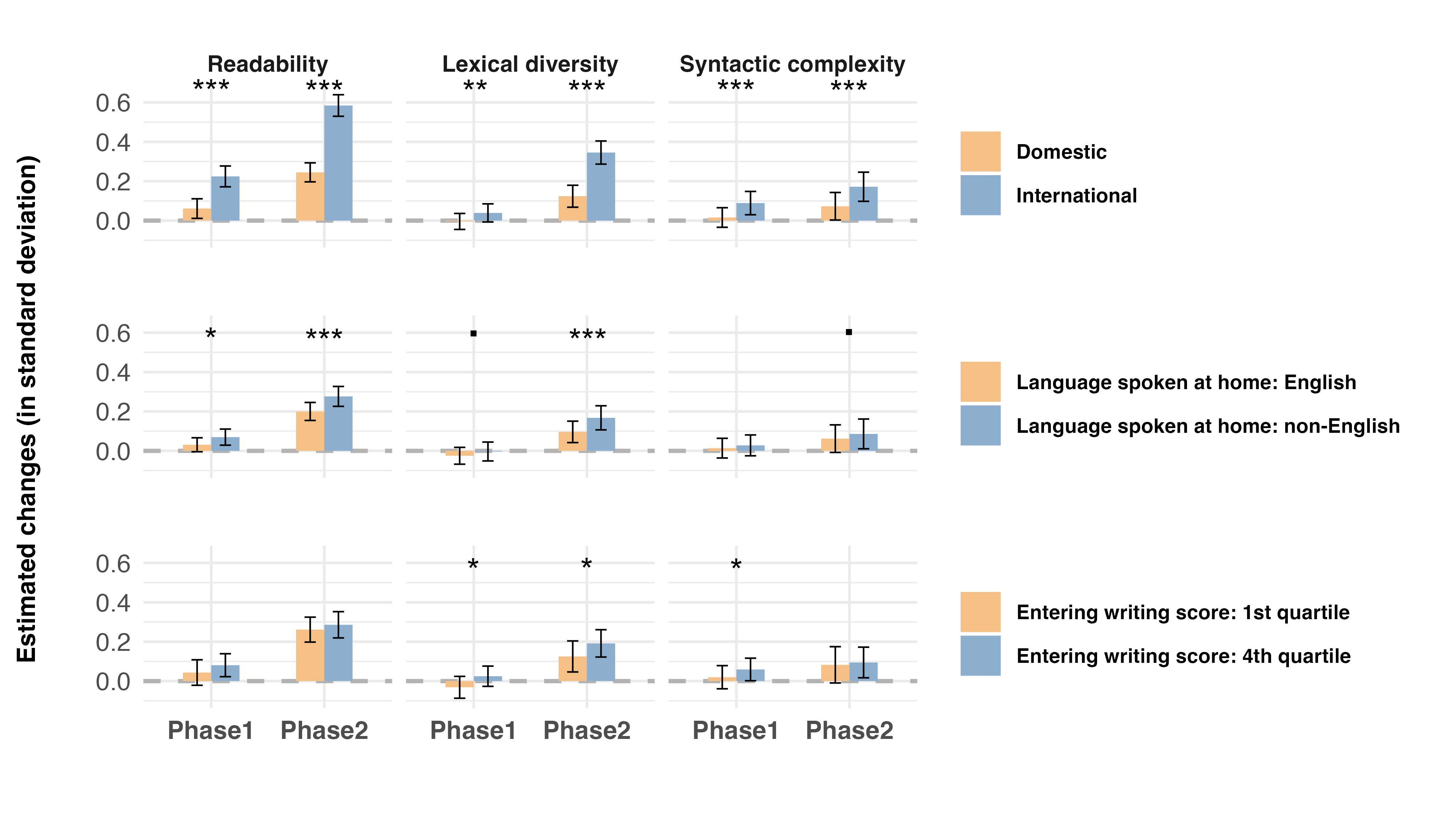}
    \caption{\textbf{Estimated changes in academic writing proficiency, by linguistic background, for writing submissions with longer content.} The sample only includes writing submissions to assignments where the median submission length was over 100 words. Each bar shows the predicted average change (in standard deviation) in a composite writing proficiency index between a post-LLM phase and the pre-LLM period for a given student group. Phase 1: January 2023 to June 2023; Phase 2: October 2023 to March 2024. Error bars indicate 90\% confidence intervals. Statistical significance for the difference between adjacent bars (interaction term between the group and the phase indicators in the regression model) is denoted by asterisks: $p<0.10(\cdot)$, $p<0.05(^*)$, $p<0.01(^{**})$, $p<0.001(^{***})$.}
\label{fig:s6}
\end{figure}

\begin{figure}[ht]
\captionsetup[subfigure]{justification=centering}
\centering
\begin{subfigure}{0.85\textwidth}
\centering
\includegraphics[width=0.85\textwidth]{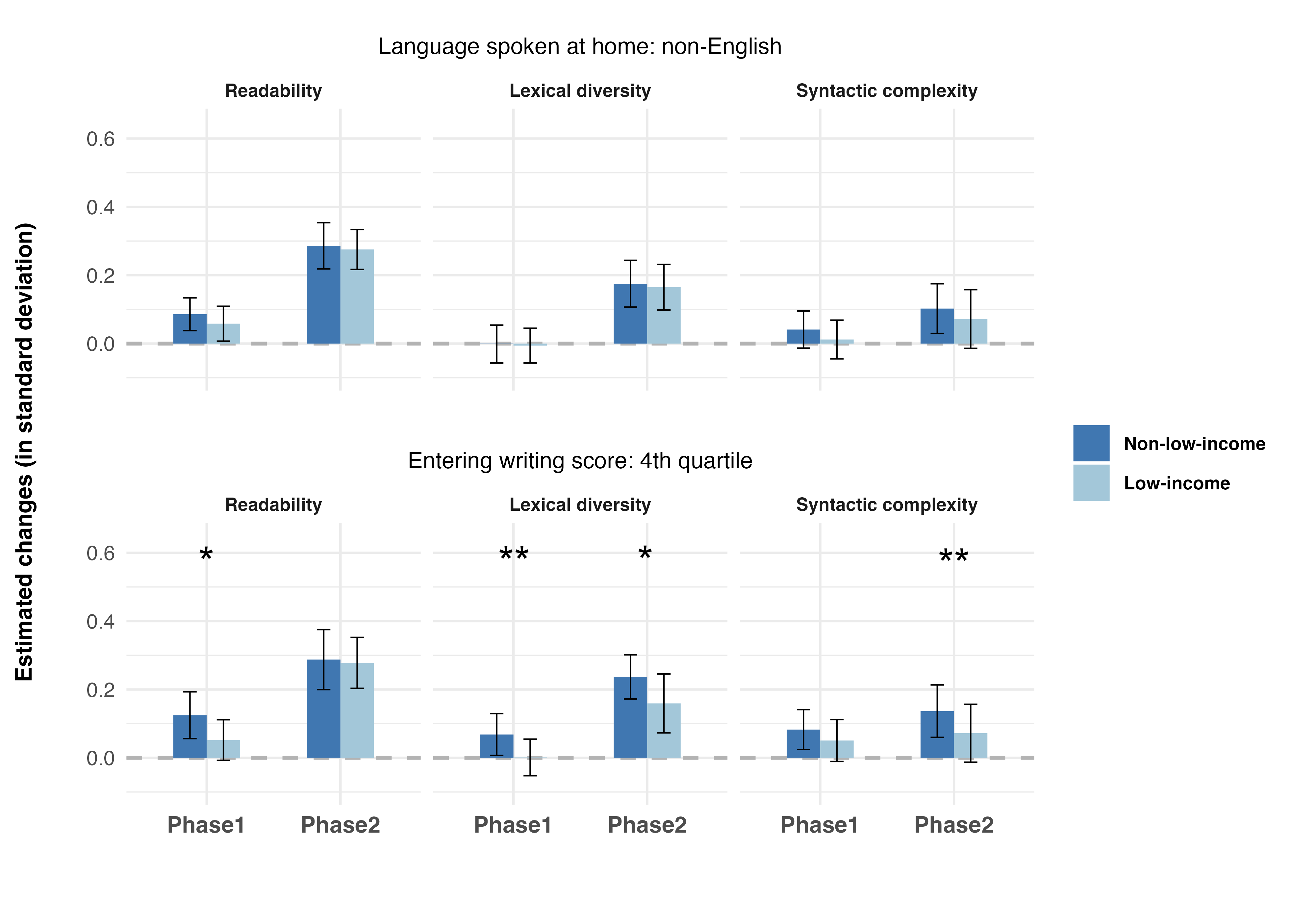}
\caption{SES: Family income}
\label{fig:s7a}
\end{subfigure}
\begin{subfigure}{0.85\textwidth}
\centering
\includegraphics[width=0.85\textwidth]{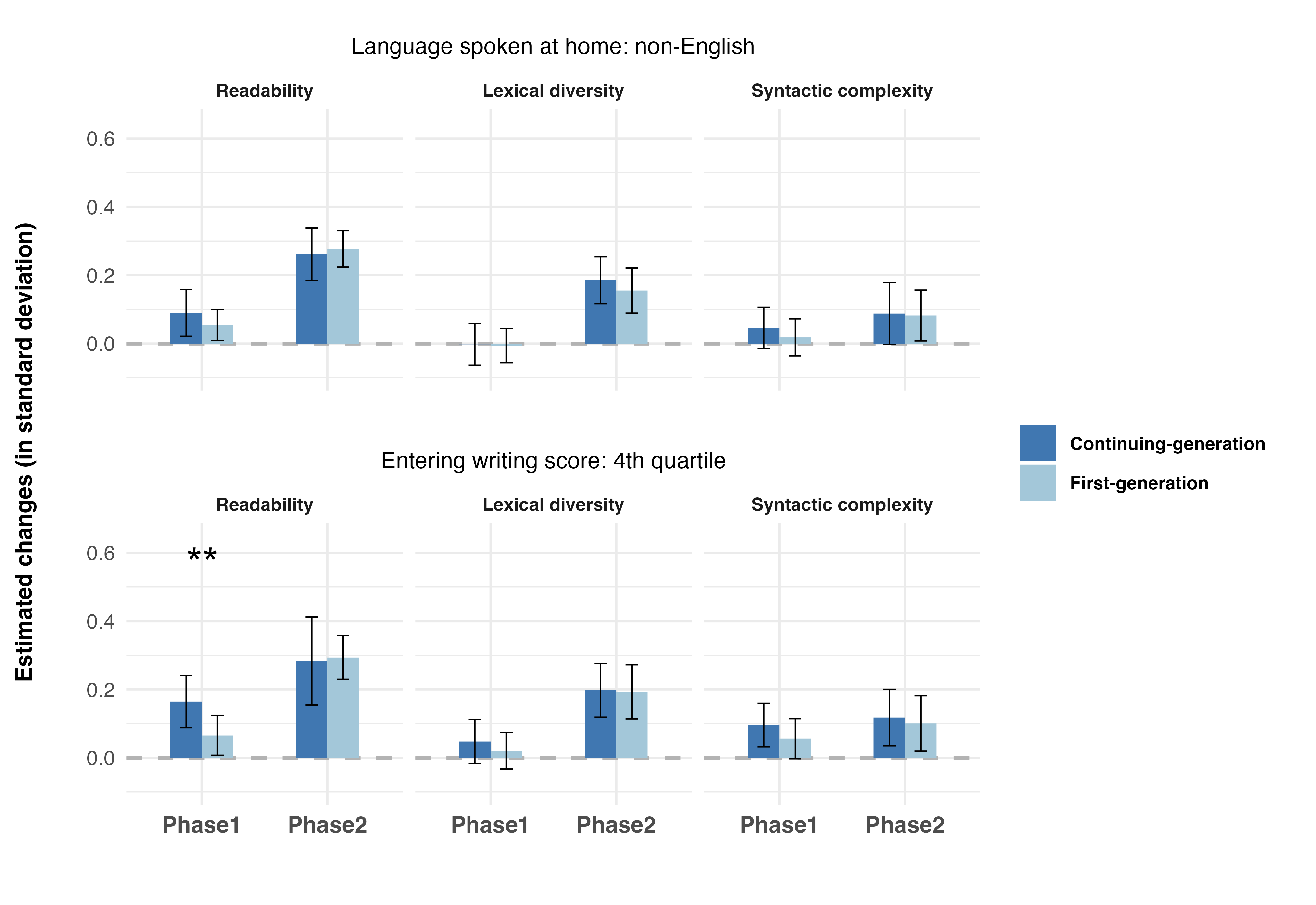}
\centering\caption{SES: Parental education}
\label{fig:s7b}
\end{subfigure}
\caption{
\small
\textbf{Estimated changes in academic writing proficiency among linguistically disadvantaged student groups, by socioeconomic status (SES), for writing submissions with longer content.} The sample only includes writing submissions to assignments where the median submission length was over 100 words. Each bar shows the predicted average change (in standard deviation) in a composite writing proficiency index between a post-LLM phase and the pre-LLM period for a given student subgroup. Phase 1: January 2023 to June 2023; Phase 2: October 2023 to March 2024. Error bars indicate 90\% confidence intervals. Statistical significance for the difference between adjacent bars (three-way interaction term between indicators of the linguistic group, socioeconomic group and phase in the regression model) is denoted by asterisks: $p<0.10(\cdot)$, $p<0.05(^*)$, $p<0.01(^{**})$, $p<0.001(^{***})$.}
\label{fig:s7}
\end{figure}

\begin{figure}[ht]
    \centering
    \includegraphics[width=0.6\textwidth]{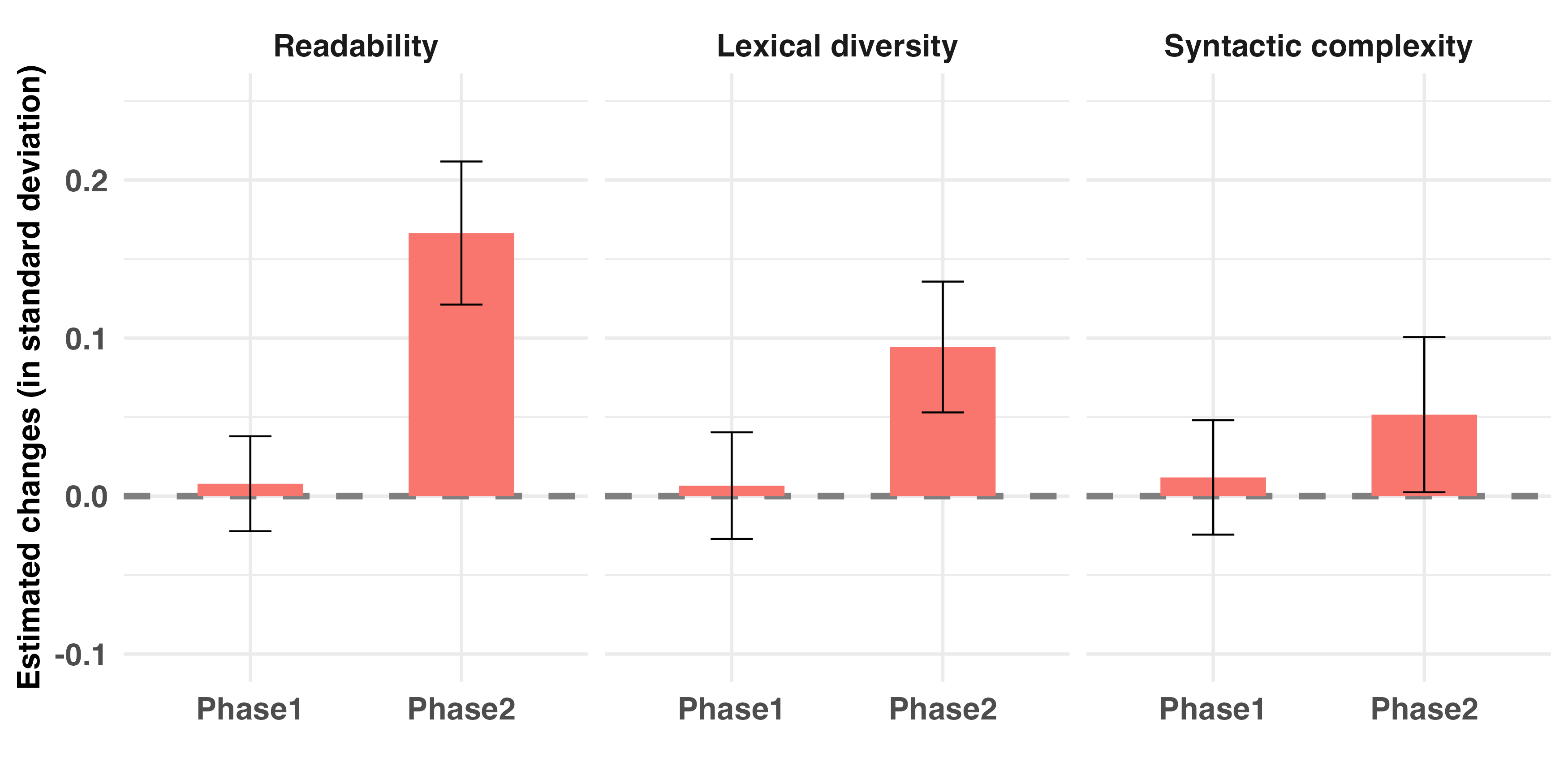}
    \caption{\textbf{Estimated changes in academic writing proficiency, based on extended time span.} The sample includes writing submissions dating back to Fall 2019 to cover academic terms affected by the COVID-19 pandemic. Each bar represents the estimate of the average change (in standard deviation) in a composite writing proficiency index between a post-LLM phase and the pre-LLM period in the data. Phase 1: January 2023 to June 2023; Phase 2: October 2023 to March 2024. Error bars indicate 90\% confidence intervals.}
\label{fig:s8}
\end{figure}

\begin{figure}[ht]
    \centering
    \includegraphics[width=0.85\textwidth]{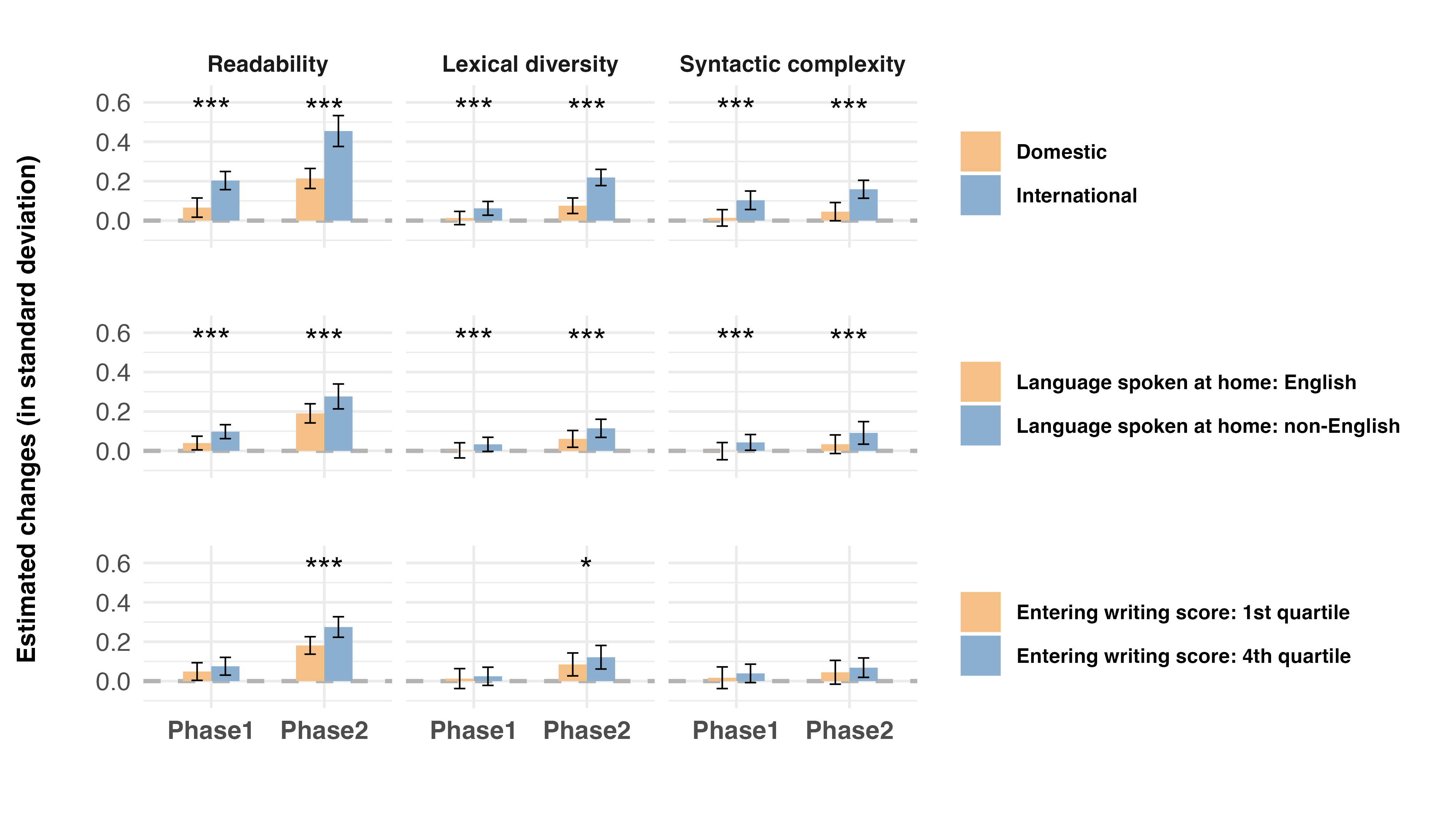}
    \caption{\textbf{Estimated changes in academic writing proficiency, by linguistic background, based on extended time span.} The sample includes writing submissions dating back to Fall 2019 to cover academic terms affected by the COVID-19 pandemic. Each bar shows the predicted average change (in standard deviation) in a composite writing proficiency index between a post-LLM phase and the pre-LLM period for a given student group. Phase 1: January 2023 to June 2023; Phase 2: October 2023 to March 2024. Error bars indicate 90\% confidence intervals. Statistical significance for the difference between adjacent bars (interaction term between the group and the phase indicators in the regression model) is denoted by asterisks: $p<0.10(\cdot)$, $p<0.05(^*)$, $p<0.01(^{**})$, $p<0.001(^{***})$.}
\label{fig:s9}
\end{figure}

\begin{figure}[ht]
\captionsetup[subfigure]{justification=centering}
\centering
\begin{subfigure}{0.85\textwidth}
\centering
\includegraphics[width=0.85\textwidth]{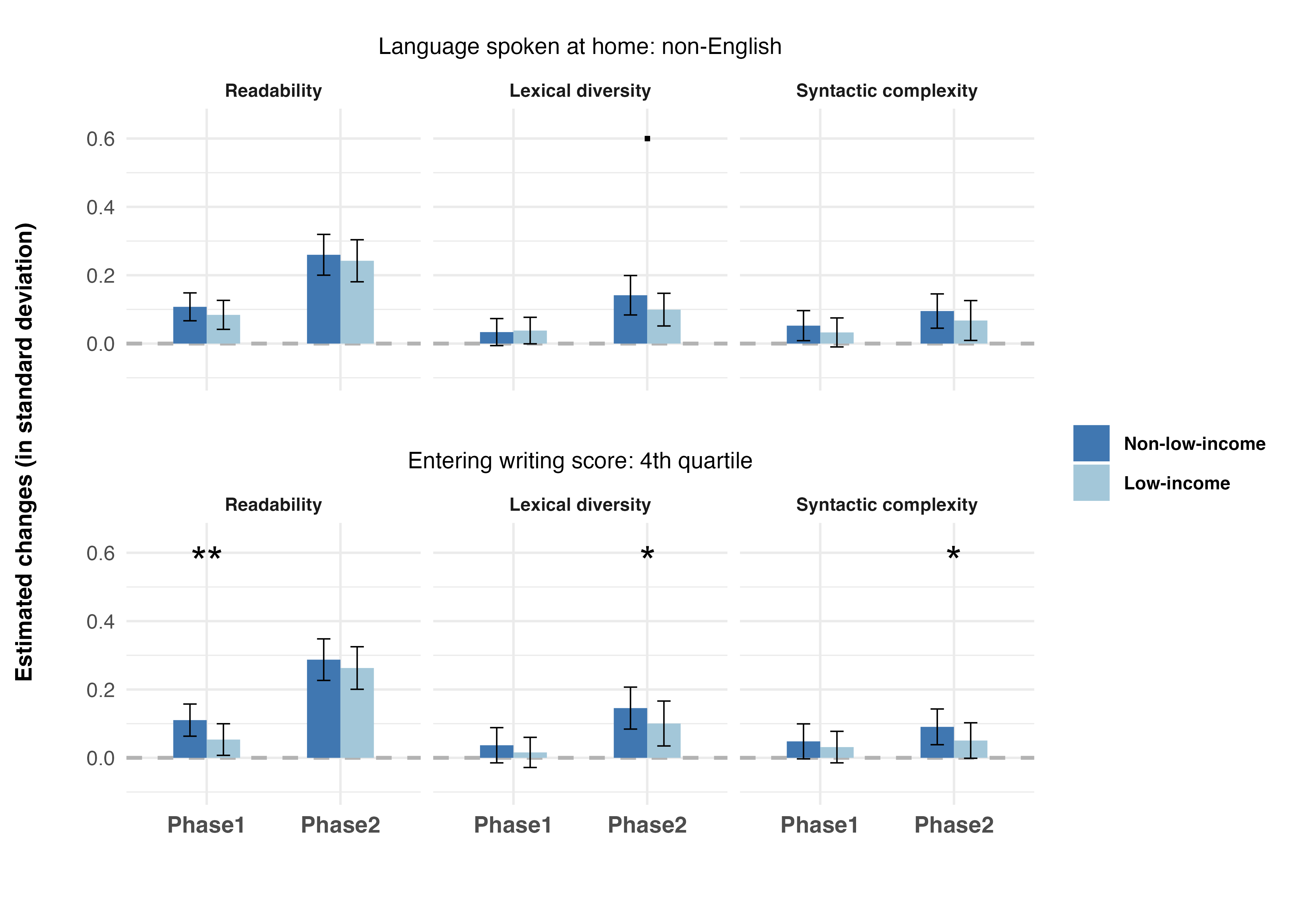}
\caption{SES: Family income}
\label{fig:s10a}
\end{subfigure}
\begin{subfigure}{0.85\textwidth}
\centering
\includegraphics[width=0.85\textwidth]{figures/rq3a_pre_covid.png}
\centering\caption{SES: Parental education}
\label{fig:s10b}
\end{subfigure}
\caption{
\small
\textbf{Estimated changes in academic writing proficiency among linguistically disadvantaged student groups, by socioeconomic status (SES), based on extended time span.} The sample includes writing submissions dating back to Fall 2019 to cover academic terms affected by the COVID-19 pandemic. Each bar shows the predicted average change (in standard deviation) in a composite writing proficiency index between a post-LLM phase and the pre-LLM period for a given student subgroup. Phase 1: January 2023 to June 2023; Phase 2: October 2023 to March 2024. Error bars indicate 90\% confidence intervals. Statistical significance for the difference between adjacent bars (three-way interaction term between indicators of the linguistic group, socioeconomic group and phase in the regression model) is denoted by asterisks: $p<0.10(\cdot)$, $p<0.05(^*)$, $p<0.01(^{**})$, $p<0.001(^{***})$.}
\label{fig:s10}
\end{figure}